\documentclass{nature}
\usepackage{epsfig}
\usepackage{amsmath}
\usepackage{amssymb,color}

\begin{document}

\title{The polarization and the fundamental sensitivity of $^{\text{39}}$K ($^{\text{133}}$Cs)-$^{\text{85}}$Rb-$^{\text{4}}$He hybrid
optical pumping spin exchange relaxation free atomic magnetometers}
\author{Jian-Hua Liu$^{1,3}$, Dong-Yang Jing$^{1,2}$, Liang-Liang Wang$^{1,2}$, Yang Li$^{4}$, Wei Quan$^{4}$, Jian-Cheng Fang$^{4}$, Wu-Ming Liu$^{\dag,1,2}$}
\maketitle

\begin{abstract}
The hybrid optical pumping spin exchange relaxation free (SERF) atomic magnetometers can realize ultrahigh sensitivity measurement of magnetic field and inertia. We have studied the $^{\text{85}}$Rb polarization of two types of hybrid optical pumping SERF magnetometers based on $^{\text{39}}$K-$^{\text{85}}$Rb-$^{\text{4}}$He and $^{\text{133}}$Cs-$^{\text{85}}$Rb-$^{\text{4}}$He respectively. Then we found that $^{\text{85}}$Rb polarization varies with the number density of buffer gas $^{\text{4}}$He and quench gas N$_{\text{2}}$, pumping rate of pump beam and cell temperature respectively, which will provide an experimental guide for the design of the magnetometer. We obtain a general formula on the fundamental sensitivity of the hybrid optical pumping SERF magnetometer due to shot-noise. The formula describes that the fundamental sensitivity of the magnetometer varies with the number density of buffer gas and quench gas, the pumping rate of pump beam, external magnetic field, cell effective radius, measurement volume, cell temperature and measurement time. We obtain a highest fundamental sensitivity of $1.5073$ $aT/Hz^{1/2}$ ($1$ $aT=10^{-18}$ $T$) with $^{\text{39}}$K-$^{\text{85}}$Rb-$^{\text{4}}$He magnetometer between above two types of magnetometers when $^{\text{85}}$Rb polarization is $0.1116$. We estimate the fundamental sensitivity limit of the hybrid optical pumping SERF magnetometer to be superior to $1.8359\times10^{-2}$ $aT/Hz^{1/2}$, which is higher than the shot-noise-limited sensitivity of $1$ $aT/Hz^{1/2}$ of K SERF atomic magnetometer.
\end{abstract}

\begin{affiliations}
\item
Beijing National Laboratory for Condensed Matter Physics, Institute of Physics, Chinese Academy of Sciences, Beijing 100190, China

$^{2}$School of Physical Sciences, University of Chinese Academy of Sciences, Beijing 100190, China

$^{3}$ School of Science, Beijing Technology and Business University, Beijing 100048, China

$^{4}$School of Instrument Science and Opto-Electronics Engineering, and Science and Technology on Inertial Laboratory, Beihang University, Beijing 100191, China

$^\dag$e-mail: wliu@iphy.ac.cn

\end{affiliations}

In recent years, ultrahigh sensitive magnetic field measurement technology has become a hotspot in research of weak magnetic field. In the field of biomedicine, it is used in magnetoencephalography (MEG) and magnetocardiography (MCG) \cite{Johnson2013,Sander2012,Wyllie2012}. In physics, it is used to analyze the magnetism of material and measure the symmetry broken of charge conjugation, parity transformation and time reversal (CPT) \cite{Brown2010,Seltzer2008,Kornack2005}. At present, the sensitivity of the spin exchange relaxation free (SERF) atomic magnetometer is the highest in the ultrahigh sensitive magnetometers\cite{Wyllie2012,Fang2013,Ledbetter2008,Nelson2001,Kominis2003}. The shot-noise limit of the K SERF magnetometer\cite{Allred2002} is estimated to be $2$ $aT/Hz^{1/2}$ and with more optimization, it should be possible to approach the shot-noise-limited sensitivity in the range $10-1$ $aT/Hz^{1/2}$ for K SERF magnetometer\cite{Kominis2003}. The effects of the spin-exchange relaxation can be suppressed in the SERF regime, when the spin-exchange rate is much larger than the Larmor precession frequency\cite{Happer1973,Happer1977}. The SERF regime can be reached by operating with sufficiently high alkali metal number density (at higher temperature) and in sufficiently low magnetic field\cite{Happer1977,Dong2012}.

It was found that hybrid optical pumping can make the SERF magnetometer realize higher experimental detecting sensitivity and more homogeneous atomic spin polarization\cite{Ito2012} and it is suitable for quantum nondestructive measurement\cite{Romalis2010}. Ito et al.\cite{Ito2011,Ito2013} realized a sensitivity of $3\times $$10^{4}$ $aT/Hz^{1/2}$ in magnetic field measurement by SERF atomic magnetometers by hybrid optical pumping of K-Rb. Fang et al.\cite{Fang2014} obtained a sensitivity of approximately $5\times $$10^{3}$ $aT/Hz^{1/2}$ by optimizing the parameters of SERF magnetometer based on K-Rb hybrid optical pumping. Li et al.\cite{LiYang2016} measured the magnetic field sensitivity better than $700$ $aT/Hz^{1/2}$ by a subfemtotesla atomic magnetometer based on hybrid optical pumping of K-Rb. However, there is almost no work about the systematic analysis of the influence factors on the polarization and the fundamental sensitivity of K (Cs)-Rb-He hybrid optical pumping SERF atomic magnetometers. We need more practical methods to obtain higher fundamental sensitivity of the hybrid optical pumping SERF atomic magnetometer.

In this report, we obtain a general formula on the fundamental sensitivity of the hybrid optical pumping SERF magnetometer, which describes the fundamental sensitivity of the magnetometer varying with the number density of buffer gas and quench gas, pumping rate of pump beam, external magnetic field, cell effective radius (the shape of the cell is roughly spherical), measurement volume, cell temperature and measurement time. We have investigated two types of hybrid optical pumping SERF atomic magnetometers based on $^{\text{39}}$K ($^{\text{133}}$Cs)-$^{\text{85}}$Rb-$^{\text{4}}$He ($^{\text{39}}$K ($^{\text{133}}$Cs)-$^{\text{85}}$Rb-$^{\text{4}}$He magnetometers), then found the fundamental sensitivity of $^{\text{133}}$Cs-$^{\text{85}}$Rb-$^{\text{4}}$He magnetometer is lower than the one of $^{\text{39}}$K-$^{\text{85}}$Rb-$^{\text{4}}$He magnetometer at the same cell temperature and in the SERF regime when the pumping rate of pump beam is bigger than about $1916$$s^{-1}$ and N$_{\text{2}}$ number density is bigger than about $1.974\times10^16 cm^{-3}$ at our chosen conditions. Optimizing the magnetometer parameters is advantageous to improve the sensitivity of the magnetometer in measuring weak magnetic field. Furthermore, we obtained a higher fundamental sensitivity of $1.8359\times10^{-2}$ $aT/Hz^{1/2}$ with $^{\text{39}}$K-$^{\text{85}}$Rb-$^{\text{4}}$He magnetometer when the polarization of $^{\text{85}}$Rb atom is $1.3174\times10^{-4}$ and the fundamental sensitivity is higher than the shot-noise-limited sensitivity of K SERF atomic magnetometer\cite{Kominis2003} in the range $10 - 1$ $aT/Hz^{1/2}$. Among $^{\text{39}}$K, $^{\text{85}}$Rb and $^{\text{133}}$Cs SERF magnetometers, there is a maximum temperature range for $^{\text{39}}$K to make the magnetometer in the SERF regime with the number density of $^{\text{39}}$K satisfies the conditions of the SERF regime, so the SERF magnetometer based on $^{\text{39}}$K is suitable for an environment with the temperature varying drastically. These findings not only optimize the parameters for the SERF regime, but also provide an experimental guide for the design of the hybrid optical pumping SERF magnetometer.

\section*{Results}

\subsection{The number density of alkali-metal atoms.}

The alkali metal vapor cell (the shape of the cell is roughly spherical) of the SERF atomic magnetometer based on hybrid optical pumping contains two types of alkali metal atoms, they are $^{\text{39}}$K-$^{\text{85}}$Rb or $^{\text{133}}$Cs-$^{\text{85}}$Rb. $^{\text{133}} $Cs can reach large saturation vapor pressure at lower temperature\cite%
{Alcock1984} and realize SERF regime at lower temperature, which has more advantages for low temperature conditions. $^{\text{39}}$K has the highest theoretical sensitivity, so we study the hybrid optical pumping SERF atomic magnetometers based on $^{\text{39}}$K-$^{\text{85}}$Rb and $^{\text{133}}$Cs-$^{\text{85}}$Rb respectively. We select $^{\text{4}}$He as the buffer gas and take N$_{\text{2}}$ as quench gas (that is $^{\text{39}}$K ($^{\text{133}}$Cs)-$^{\text{85}}$Rb-$^{\text{4}} $He magnetometers). He gas suppresses the spin relaxation caused by wall collisions, colliding with excited alkali metal atoms and absorbing the energy, N$_{\text{2}}$ gas restrains radiative deexcitation of the excited alkali metal atoms\cite{Ito2016}. One type of alkali-metal atom which is directly pumped and polarized by a circularly polarized pump beam is called A and the other type of alkali-metal atom which is polarized by the spin-exchange collisions with A is called B in the hybrid optical pumping SERF magnetometer\cite{Romalis2010,Jau2004}, we take $^{\text{39}}$K or $^{\text{133}}$Cs as A respectively, select $^{\text{85}}$Rb as B in the SERF regime. The number density of alkali-metal vapor and the polarization of alkali-metal vapor are two of the most important parameters of the cell\cite{Shang2016}.

The saturated density of the alkali-metal atoms vapor in units of $cm^{-3}$ at cell temperature $T$ in Kelvin is given by the following equation\cite{Seltzer2008}
\begin{equation}
n_{sat}=\frac{1}{T}10^{21.866+A_{1}-B_{1}/T},  \label{1}
\end{equation}%
where $n_{sat}$ is the saturated density of alkali-metal ($^{\text{39}}$K, $^{\text{85}}$Rb and $^{\text{133}}$Cs) atom vapor. When the alkali-metal vapor cell only contains one type of alkali-metal atom (single alkali-metal vapor cell), the number density of the alkali-metal atom equals to the saturated density of alkali-metal, the parameters $A_{1}$ and $B_{1}$ are phase parameters\cite{Alcock1984}, where $A_{1}^{K}=4.402$, $A_{1}^{Rb}=4.312$, $A_{1}^{Cs}=4.165$, $B_{1}^{K}=4453$, $B_{1}^{Rb}=4040$ and $B_{1}^{Cs}=3830$ for the temperature is higher than $400 K$.

We can obtain the number density of $^{\text{39}}$K, $^{\text{133}}$Cs and $^{\text{85}}$Rb varying with the cell temperature for the single alkali-metal vapor cell from equation (1) as shown in Fig. 1. When the number density of $^{\text{39}}$K, $^{\text{85}}$Rb and $^{\text{133}}$Cs atom are the same, $^{\text{39}}$K need the highest temperature. In general, the number density of the alkali metal atoms is $10^{13}$ $cm^{-3}$ to $10^{14}$ $cm^{-3}$ in the SERF regime, we can find that there is a maximum temperature range for $^{\text{39}}$K to make the magnetometer in the SERF regime with the number density of $^{\text{39}}$K satisfies the conditions of the SERF regime, therefore the SERF magnetometer based on $^{\text{39}}$K is suitable for an environment with the temperature varying drastically. If we increase the cell temperature, we can obtain higher number density of alkali-metal atoms, however, the cell glass will be corroded, the laser power and heating equipment will be unable to bear and there will be other problem experimentally when the number density of alkali-metal atoms is greater than or equal to $10^{15}$ $cm^{-3}$. What's more, the optical depth will be too big so that the laser will be largely absorbed by the atoms. If the vapor cell is made of special glass, the laser power is very big and the volume of the vapor cell is very small, we can appropriately increase the temperature of the vapor cell. Depending on equation (1), when cell temperature $T=457.5$ $K$ for single alkali-metal vapor cell, we obtain the number density of $^{\text{39}}$K is $n_{K}=7.4864\times $$10^{13}$ $cm^{-3}$, the number density of $^{\text{85}}$Rb is $n_{Rb}=9.9776\times $$10^{14}$ $cm^{-3}$, the number density of $^{\text{133}}$Cs is $n_{Cs}=4.8642\times $$10^{14}$ $cm^{-3}$. When $T=457.6$ $K$, we obtain $n_{K}=7.5216\times $$10^{13}$ $cm^{-3}$, $n_{Rb}=1.0017\times $$10^{15}$ $cm^{-3}$, $n_{Cs}=4.8848\times $$10^{14}$ $cm^{-3}$. Because SERF regime can be reached by operating with sufficiently high alkali metal number density (at higher temperature) and in sufficiently low magnetic field\cite{Happer1977,Dong2012}, we choose $T=457.5$ $K$ as the highest temperature to reduce the corrosion of alkali metal atoms to the vapor cell and make the magnetometer be in the SERF regime.

\subsection{The polarization of alkali-metal atom.}

Considering the spin-exchange between two types of alkali-metal atoms A and B in the hybrid vapor cell, here, we assume that the vapor densities obey the Raoult's law\cite{Babcock2003}, $n_{B}\approx f_{B}n_{sat}^{B}$, where $f_{B}$ is the mole fraction of atom B in the metal and $n_{sat}^{B}$ is the saturated vapor density for pure atom B metal. When the mole fraction of atom B is $0.97$, we can obtain the number density of atom A and B, $n_{A}\approx 0.03n_{sat}^{A}$, $n_{B}\approx 0.97n_{sat}^{B}$. The spin polarization $P$ of each type of atoms in zero magnetic field can be described as\cite{Fang2014},
\begin{equation}
P_{A}=\frac{R_{SE}^{B}P_{B}+R_{p}}{R_{p}+R_{SD}^{A}+R_{SE}^{B}}\text{,}  \label{2}
\end{equation}%
\begin{equation}
P_{B}=\frac{R_{SE}^{A}P_{A}}{R_{SD}^{B}+R_{SE}^{A}}\text{,}  \label{3}
\end{equation}%
\begin{eqnarray}
R _{SD}^{A} &=&n_{A}\sigma _{SD}^{A-A}\bar{v}_{A-A}+n_{He}\sigma _{SD}^{A-He}\bar{v}_{A-He}+n_{N_{2}}\sigma _{SD}^{A-N_{2}}\bar{v}_{A-N_{2}}+n_{A}\sigma
_{SD}^{A-B}\bar{v}_{A-B}\text{,}  \label{4} \\
R_{SD}^{B} &=&n_{B}\sigma _{SD}^{B-B}\bar{v}_{B-B}+n_{He}\sigma _{SD}^{B-He}\bar{v}_{B-He}+n_{N_{2}}\sigma _{SD}^{B-N_{2}}\bar{v}_{B-N_{2}}+n_{B}\sigma
_{SD}^{B-A}\bar{v}_{B-A}\text{,}  \label{5}
\end{eqnarray}%
where A represents alkali-metal atom $^{\text{39}}$K or $^{\text{133}}$Cs, B represents alkali-metal atom $^{\text{85}}$Rb. $R_{SE}^{A}$ is the spin-exchange rate of atom B with atom A, $R_{SE}^{A}=k_{SE}n_{A}=n_{A}\sigma _{SE}^{B-A}\bar{v}_{B-A}$, $R_{SE}^{B}$ is the spin-exchange rate of atom A with atom B, $R_{SE}^{B}=k_{SE}n_{B}=n_{B}\sigma _{SE}^{A-B}\bar{v}_{A-B}$, $R_{p}$ is the pumping rate of pump beam, which is mainly determined by pumping laser parameters\cite{Kornack2005}, $R_{B}=k_{SE}n_{A}P_{A}$ is the pumping rate of atom B\cite{Ito2011}, $R_{SD}$ is the spin destruction relaxation rate, $k_{SE}$ is the spin-exchange rate constant, $n$ is the number density of atoms, $\bar{v}_{alkali-alkali}$ is the relative velocity between the alkali atoms, $\bar{v}_{alkali-alkali}=\sqrt{\frac{8\kappa _{B}T}{\pi m_{alkali-alkali}}}$, $\kappa _{B}$ is Boltzmann's constant, $T$ is cell temperature, $\bar{v}_{alkali-N_{2}}=\sqrt{\frac{8\kappa _{B}T}{\pi m_{alkali-N_{2}}}}$ is the relative velocity between the alkali atoms and quench gas N$_{\text{2}}$ respectively, the reduced mass of alkali atoms and quench gas N$_{\text{2}}$ is $m_{alkali-N_{2}}=\frac{%
m_{alkali}m_{N_{2}}}{m_{alkali}+m_{N_{2}}}$, the relative velocity between the alkali atoms and buffer gas $^{\text{4}}$He is $\bar{v}_{alkali-He}=\sqrt{%
\frac{8\kappa _{B}T}{\pi m_{alkali-He}}}$, the reduced mass of alkali atoms and buffer gas $^{\text{4}}$He is $m_{alkali-He}=\frac{m_{alkali}m_{He}}{m_{alkali}+m_{He}}$, $k_{SE}$ for different alkali metal atoms are nearly the same\cite{Gibbs1967,Ressler1969,Shao2005,Ghosh2010,Quan2015}, $k_{SE}=\left\langle \sigma _{SE}v\right\rangle $, $k_{SE}^{Rb-Cs}= k_{SE}^{Cs-Rb}$, $k_{SE}^{Rb-K}= k_{SE}^{K-Rb}$, the spin-exchange cross section of $^{\text{39}}$K, $^{\text{85}}$Rb and $^{\text{133}}$Cs is $\sigma _{SE}^{K}=1.8\times 10^{-14}$ $cm^{2}$, $\sigma_{SE}^{Rb}=1.9\times 10^{-14}$ $cm^{2}$ and $\sigma _{SE}^{Cs}=2.1\times10^{-14}$ $cm^{2}$ respectively\cite{Ressler1969,Aleksandrov1999}, $k_{SE}^{Cs-Rb}=\sigma _{SE}^{Cs-Rb}\sqrt{\frac{m_{Rb}+m_{Cs}}{m_{Rb}m_{Cs}}}\sqrt{\frac{8\kappa _{B}T}{\pi }}$ and $k_{SE}^{K-Rb}=\sigma _{SE}^{K-Rb}\sqrt{\frac{m_{Rb}+m_{K}}{m_{Rb}m_{K}}}\sqrt{\frac{8\kappa _{B}T}{\pi }}$, where $\sigma _{SE}^{K-Rb}$ is the spin-exchange cross section of $^{\text{39}}$K and $^{\text{85}}$Rb by spin-exchange collisions with each other, $\sigma _{SE}^{Cs-Rb}$ is the spin-exchange cross section of $^{\text{133}}$Cs and $^{\text{85}}$Rb by spin-exchange collisions with each other, for $\sigma _{SE}^{^{\text{87}}Rb-^{\text{87}}Rb}=(1.9\pm0.2)\times10^{-14}$ $cm^{2}$, $\sigma _{SE}^{^{\text{87}}Rb-^{\text{133}}Cs}= (2.3\pm0.2)\times10^{-14}$ $cm^{2}$, however, we don't find the values of $\sigma_{SE}^{^{\text{39}}K-^{\text{85}}Rb}$ and $\sigma_{SE}^{^{\text{133}}Cs-^{\text{85}}Rb}$, we take $\sigma_{SE}^{^{\text{39}}K-^{\text{85}}Rb}=\sigma _{SE}^{^{\text{85}}Rb-^{\text{39}}K}\approx \sigma _{SE}^{K}\frac{n_{K}}{n_{K}+n_{Rb}}+\sigma _{SE}^{Rb}\frac{n_{Rb}}{n_{K}+n_{Rb}}$, $\sigma_{SE}^{^{\text{133}}Cs-^{\text{85}}Rb}=\sigma _{SE}^{^{\text{85}}Rb-^{\text{133}}Cs}\approx \sigma _{SE}^{Cs}\frac{n_{Cs}}{n_{Cs}+n_{Rb}}+\sigma _{SE}^{Rb}\frac{n_{Rb}}{n_{Cs}+n_{Rb}}$  by considering the weight of $^{\text{39}}$K, $^{\text{85}}$Rb and $^{\text{133}}$Cs.

The spin-destruction cross section of $^{\text{39}}$K-$^{\text{39}}$K, $^{\text{85}}$Rb-$^{\text{85}}$Rb and $^{\text{133}}$Cs-$^{\text{133}}$Cs are $\sigma _{SD}^{K}=1\times 10^{-18}$ $cm^{2}$, $\sigma_{SD}^{Rb}=1.6\times 10^{-17}$ $cm^{2}$ and $\sigma _{SD}^{Cs}=2\times10^{-16}$ $cm^{2}$ respectively\cite{Vliegen2001,Walker1997,Bhaskar1980}, $\sigma _{SD}^{K-Rb}$, $\sigma _{SD}^{Rb-K}$, $\sigma_{SD}^{Cs-Rb}$ and $\sigma_{SD}^{Rb-Cs}$ are the spin-destruction cross section of $^{\text{39}}$K and $^{\text{85}}$Rb, $^{\text{85}}$Rb and $^{\text{39}}$K, $^{\text{133}}$Cs and $^{\text{85}}$Rb, $^{\text{85}}$Rb and $^{\text{133}}$Cs by collisions with each other respectively. However, we do not find the values of $\sigma _{SD}^{K-Rb}$, $\sigma _{SD}^{Rb-K}$, $\sigma_{SD}^{Cs-Rb}$ and $\sigma_{SD}^{Rb-Cs}$ in any references, depending on the spin-destruction cross section of $^{\text{39}}$K-$^{\text{39}}$K, $^{\text{85}}$Rb-$^{\text{85}}$Rb and $^{\text{133}}$Cs-$^{\text{133}}$Cs, we take $\sigma_{SD}^{K-Rb}=\sigma _{SD}^{Rb-K}\approx \sigma _{SD}^{K}\frac{n_{K}}{n_{K}+n_{Rb}}+\sigma _{SD}^{Rb}\frac{n_{Rb}}{n_{K}+n_{Rb}}$, $\sigma_{SD}^{Cs-Rb}=\sigma _{SD}^{Rb-Cs}\approx \sigma _{SD}^{Cs}\frac{n_{Cs}}{n_{Cs}+n_{Rb}}+\sigma _{SD}^{Rb}\frac{n_{Rb}}{n_{Cs}+n_{Rb}}$ by considering the weight of $^{\text{39}}$K, $^{\text{85}}$Rb and $^{\text{133}}$Cs. The spin-destruction cross section of $^{\text{4}}$He and $^{\text{39}}$K, $^{\text{4}}$He and $^{\text{85}}$Rb, $^{\text{4}}$He and $^{\text{133}}$Cs are $\sigma _{SD}^{K-He}=8\times 10^{-25}$ $cm^{2}$, $\sigma _{SD}^{Rb-He}=9\times10^{-24}$ $cm^{2}$ and $\sigma _{SD}^{Cs-He}=2.8\times 10^{-23}$ $cm^{2}$ respectively\cite{Allred2002}. The spin-destruction cross section of N$_{\text{2}}$ and $^{\text{39}}$K, N$_{\text{2}}$ and $^{\text{85}}$Rb, N$_{\text{2}}$ and $^{\text{133}}$Cs is $\sigma _{SD}^{K-N_{2}}=7.9\times10^{-23}$ $cm^{2}$, $\sigma _{SD}^{Rb-N_{2}}=1\times 10^{-22}$ $cm^{2}$ and $%
\sigma _{SD}^{Cs-N_{2}}=5.5\times 10^{-22}$ $cm^{2}$ respectively\cite{Kadlecek1998,Allred2002}.

Substitute equation (2) into equation (3), we obtain
\begin{equation}
P_{B}=\frac{R_{p}R_{SE}^{A}}{(R_{SD}^{B}+R_{SE}^{A})(R_{p}+R_{SD}^{A}+R_{SE}^{B})-R_{SE}^{A}R_{SE}^{B}}\text{,}  \label{6}
\end{equation}

We take one of $^{\text{4}}$He number density $n_{He}$, N$_{\text{2}}$ number density $n_{N2}$, cell temperature $T$ and pumping rate of pump beam $R_{p}$ by equation (6) as a variable (other parameters are invariable) to obtain the results that the polarization of the hybrid optical pumping SERF magnetometer based on $^{\text{39}}$K ($^{\text{133}}$Cs)-$^{\text{85}}$Rb-$^{\text{4}}$He respectively vary with the variable. Depending on suggestions and the typical conditions of the experiment group\cite{Fang2014,Quan2015,Chen2016S}, in order to facilitate the theoretical analysis, we take the mole fraction of atom B $f_{B}=0.97$, $n_{He}=10^{19}$ $cm^{-3}$, $n_{N2}=2\times10^{17}$ $cm^{-3}$, $R_{p}=R_{p}^{K}=R_{p}^{Cs}=200000$ $s^{-1}$ and $T=457.5$ $K$, at the moment, $^{\text{39}}$K, $^{\text{85}}$Rb and $^{\text{133}}$Cs are in the SERF regime.

The number density of $^{\text{39}}$K, $^{\text{85}}$Rb and $^{\text{133}}$Cs vary with cell temperature and the mole fraction for the hybrid alkali-metal vapor cell, so the relation among the polarization of $^{\text{85}}$Rb, the number density of alkali metal atom ($^{\text{39}}$K, $^{\text{85}}$Rb or $^{\text{133}}$Cs) and cell temperature is nonlinear in groups $^{\text{39}}$K-$^{\text{85}}$Rb and $^{\text{133}}$Cs-$^{\text{85}}$Rb when the mole fraction of the alkali-metal atoms are fixed. At the same cell temperature, the number density of $^{\text{39}}$K, $^{\text{85}}$Rb and $^{\text{133}}$Cs are different. When the number density of $^{\text{39}}$K, $^{\text{85}}$Rb and $^{\text{133}}$Cs are equivalent, the cell temperature of $^{\text{39}}$K-$^{\text{85}}$Rb and $^{\text{133}}$Cs-$^{\text{85}}$Rb magnetometers are different, and the $^{\text{85}}$Rb polarization of $^{\text{39}}$K-$^{\text{85}}$Rb and $^{\text{133}}$Cs-$^{\text{85}}$Rb magnetometers are also different.

Fig. 2 demonstrates the $^{\text{85}}$Rb polarization of $^{\text{39}}$K ($^{\text{133}}$Cs)-$^{\text{85}}$Rb-$^{\text{4}}$He magnetometers as a function of $^{\text{39}}$K, $^{\text{133}}$Cs, $^{\text{85}}$Rb, $^{\text{4}}$He, and N$_{\text{2}}$ number density, the pumping rate of pump beam and cell temperature, including the effects of spin exchange due to $^{\text{39}}$K-$^{\text{85}}$Rb, $^{\text{85}}$Rb-$^{\text{39}}$K, $^{\text{133}}$Cs-$^{\text{85}}$Rb, $^{\text{85}}$Rb-$^{\text{133}}$Cs and spin relaxation due to $^{\text{39}}$K-$^{\text{39}}$K, $^{\text{39}}$K-$^{\text{85}}$Rb, $^{\text{85}}$Rb-$^{\text{85}}$Rb, $^{\text{133}}$Cs-$^{\text{85}}$Rb, $^{\text{133}}$Cs-$^{\text{133}}$Cs, $^{\text{39}}$K-$^{\text{4}}$He, $^{\text{85}}$Rb-$^{\text{4}}$He, $^{\text{133}}$Cs-$^{\text{4}}$He collisons, $^{\text{39}}$K-N$_{\text{2}}$, $^{\text{85}}$Rb-N$_{\text{2}}$, $^{\text{133}}$Cs-N$_{\text{2}}$ destructions. The $^{\text{85}}$Rb polarization almost does not vary with $^{\text{4}}$He number density when $^{\text{4}}$He number density is smaller than a critical value about $10^{20}$ $cm^{-3}$ in $^{\text{39}}$K -$^{\text{85}}$Rb-$^{\text{4}}$He (black line in squares) and $^{\text{133}}$Cs-$^{\text{85}}$Rb-$^{\text{4}}$He (red line in dots) magnetometers, otherwise, the $^{\text{85}}$Rb polarization decreases rapidly in Fig. 2(a). The $^{\text{85}}$Rb polarization almost does not vary with N$_{\text{2}}$ number density when N$_{\text{2}}$ number density is smaller than about $2\times 10^{19}$ $cm^{-3}$ for $^{\text{39}}$K-$^{\text{85}}$Rb-$^{\text{4}}$He magnetometer and $^{\text{133}}$Cs-$^{\text{85}}$Rb-$^{\text{4}}$He magnetometer. Otherwise, the $^{\text{85}}$Rb polarization decreases rapidly in Fig. 2(b). The $^{\text{85}}$Rb polarization increases with the increasing pumping rate of pump beam $R_{p}^{K}$ and $R_{p}^{Cs}$ respectively in Fig. 2(c). The polarization of $^{\text{85}}$Rb decreases with the cell temperature increasing in Fig. 2(d). The $^{\text{85}}$Rb polarization of $^{\text{133}}$Cs-$^{\text{85}}$Rb-$^{\text{4}}$He magnetometer is bigger than the one of $^{\text{39}}$K-$^{\text{85}}$Rb-$^{\text{4}}$He magnetometer in (a)-(d).

\subsection{The fundamental sensitivity of the hybrid optical pumping SERF
atomic magnetometer.}

To improve the practicability of the hybrid optical pumping SERF atomic magnetometer, it is necessary for us to investigate the fundamental sensitivity of the magnetometer to improve the sensitivity and stability of the magnetometer and realize the miniaturization of the magnetometer. The fundamental, shot-noise-limited sensitivity of an atomic magnetometer is given by\cite{Budker2002}
\begin{equation}
\delta B=\frac{1}{\gamma\sqrt{nT_{2}Vt}}\text{,}  \label{7}
\end{equation}%
it is also the ultimate sensibility of the atomic magnetometer\cite{Allred2002}, where $n$ is the number density of alkali-metal atoms\cite{Fang2012}, $\gamma$ is their gyromagnetic ratio and the effective $\gamma$ for sensitivity estimates is $\gamma=\frac{g\mu _{B}}{\hbar }$ (equation (7) of Ref. 11) in our magnetometer operating at zero field, we replace it by electron gyromagnetic ratio $\gamma _{e}$ ($\gamma _{e}=\frac{g\mu _{B}}{\hbar }$)\cite{Fang2012,Seltzer2008}, $g$ is the electron $g$-factor, $\mu _{B}$ is the Bohr magneton, $V$ is the measurement volume, $t$ is the measurement time, $T_{2}$ is the transverse spin relaxation time\cite{Ledbetter2008}, $\frac{1}{T_{2}}=R _{SD}+R_{wall}+R_{SE}^{ee}$. For the transverse spin relaxation time of the hybrid optical pumping SERF atomic magnetometer, we need consider the spin destruction relaxation $R _{SD}$ caused by He, N$_{\text{2}}$, alkali metal atom A and B, the relaxation rates due to diffusion of alkali metal atoms A and B to the wall\cite{Kornack2005} $R_{wall}^{A}$ and $R_{wall}^{B}$, the relaxation rate due to alkali-alkali spin-exchange collisions\cite{Fang2015} $R_{SE}^{ee}$, $R_{SE}^{ee}=R_{SE}^{AA}+R_{SE}^{AB}+R_{SE}^{BA}+R_{SE}^{BB}$, which cannot be ignored for large external magnetic field $B$ and is negligible in SERF regime (when $T$ is higher than $418.3$ $K$, $B$ is smaller than $10^{-10}$ $T$, $R_{SE}^{ee}\approx 0$), the pumping rate of pump beam $R_{p}$ and the pumping rate of atom B $R_{B}$ ($R_{B}$ is a function of $R_{p}$), buffer gas is $^{\text{4}}$He, quench gas is N$_{\text{2}}$, therefore, $\frac{1}{T_{2}}=R _{SD}^{A}+R _{SD}^{B}+R_{wall}^{A}+R_{wall}^{B}+R_{SE}^{AA}+R_{SE}^{AB}+R_{SE}^{BA}+R_{SE}^{BB}+R_{p}+R_{B}$, we substitute this term into equation (7) and obtain
\begin{equation}
\delta B=\frac{\sqrt{R _{SD}^{A}+R _{SD}^{B}+R_{wall}^{A}+R_{wall}^{B}+R_{SE}^{AA}+R_{SE}^{AB}+R_{SE}^{BA}+R_{SE}^{BB}+R_{p}+R_{B}}}{\gamma _{e}\sqrt{nVt}}\text{.}  \label{8}
\end{equation}
However, because alkali metal atom B is probed atom, only these items associated with atom B will be considered in the experiments, we don't consider those items irrelevant to atom B and acquire the fundamental sensitivity of the hybrid optical pumping SERF atomic magnetometer due to the shot-noise as following
\begin{equation}
\delta B'=\frac{\sqrt{R_{wall}^{B}+R_{SD}+R_{B}+R_{SE}^{BB}+R_{SE}^{AB}+R_{SE}^{BA}}}{\gamma _{e}\sqrt{n_{B}Vt}}\text{.}  \label{9}
\end{equation}
where $R_{wall}=q(P)D_{buffer}^{alkali}\left( \frac{\sqrt{1+T/273.15}}{P_{buffer}/1amg}\right) \left( \frac{\pi }{a}\right) ^{2}+q(P)D_{quench}^{alkali}\left( \frac{\sqrt{1+T/273.15}}{P_{quench}/1amg}\right) \left( \frac{\pi }{a}\right) ^{2}$, the second term that alkali-metal atoms diffuse in the quench gas sometimes is ignored in the experiment, but we find that the harmonic mean of the diffusion coefficients in He and N$_{\text{2}}$ are used for $D$ (diffusion constant of the alkali atom within the gas) in the calculations of Ref. 22, hence we also consider the second term, $R_{wall}^{B}=R_{wall}^{B-He}+R_{wall}^{B-N2}$. $q(P)$ is the nuclear slowing-down factor of alkali-metal atom\cite{Appelt1998}, $q(P)_{K}=\frac{6+2P^{2}}{1+P^{2}}$ for $^{39}K$\ atom, $q(P)_{Rb}=\frac{38+52P^{2}+6P^{4}}{3+10P^{2}+3P^{4}}$ for $^{85}Rb$\ atom, $q(P)_{Cs}=\frac{22+70P^{2}+34P^{4}+2P^{6}}{1+7P^{2}+7P^{4}+P^{6}}$ for $^{133}Cs$\ atom, $D_{buffer}^{alkali}$ is the diffusion constant of the alkali atom within the buffer gas\cite{Franz1982,Franz1976,Franz1974} in units of $cm^{2}/s$ and is given at $1$ $amg$ and $273$ $K$, $D_{quench}^{alkali}$ is the diffusion constant of the alkali atom within the quench gas\cite{Silver1984,Franz1976,Franz1974} in units of $cm^{2}/s$ and is given at $1$ $amg$ and $273$ $K$, $1amg=2.69\times 10^{19}$ $cm^{-3}$, $D_{He}^{K}=0.35$ $cm^{2}/s$, $D_{He}^{Rb}=0.5$ $cm^{2}/s$, $D_{He}^{Cs}=0.29$ $cm^{2}/s$, $D_{N2}^{K}=0.2$ $cm^{2}/s$, $D_{N2}^{Rb}=0.19$ $cm^{2}/s$, $D_{N2}^{Cs}=0.098$ $cm^{2}/s$, $P_{buffer}$ is the pressure intensity of buffer gas in $amg$, $P_{quench}$ is the pressure intensity of quench gas in $amg$, $a$ is the equivalent radius of vapor cell, $R _{SD} =n_{He}\sigma_{SD}^{B-He}\bar{v}_{B-He}+n_{N_{2}}\sigma_{SD}^{B-N_{2}}\bar{v}_{B-N_{2}}+n_{B}\sigma _{SD}^{B-B}\bar{v}_{B-B}+n_{B}\sigma_{SD}^{B-A}\bar{v}_{B-A}+n_{A}\sigma_{SD}^{A-B}\bar{v}_{A-B}$, $R_{SE}^{BB}=\left( \frac{g\mu _{B}B}{q(0)_{B}\hbar }\right) ^{2}\frac{q(0)_{B}^{2}-(2I_{B}+1)^{2}}{2k_{SE}^{B-B}n_{B}}$, $B$ is the external magnetic field, $q(0)$ is the low polarization limit of the slowing-down factor, $I$ is nuclear spin of the alkali-metal atoms, for $^{\text{39}}$K, $^{\text{85}}$Rb and $^{\text{133}}$Cs, $I_{K}=1.5$, $I_{Rb}=2.5$, $I_{Cs}=3.5$, the relaxation rate due to alkali-alkali spin-exchange collisions $R_{SE}^{BB}=\left( \frac{g\mu _{B}B}{q(0)_{B}\hbar }\right) ^{2}\frac{q(0)_{B}^{2}-(2I_{B}+1)^{2}}{2k_{SE}^{B-B}n_{B}}$, $R_{SE}^{BA}=\left( \frac{g\mu _{B}B}{q(0)_{A}\hbar }\right) ^{2}\frac{q(0)_{A}^{2}-(2I_{A}+1)^{2}}{2k_{SE}^{B-A}n_{A}}$, $R_{SE}^{AB}=\left( \frac{g\mu _{B}B}{q(0)_{B}\hbar }\right) ^{2}\frac{q(0)_{B}^{2}-(2I_{B}+1)^{2}}{2k_{SE}^{A-B}n_{B}}$. With sufficiently high alkali metal number density (at higher temperature) and in sufficiently low magnetic field, $R_{SE}^{BB}\approx R_{SE}^{BA}\approx R_{SE}^{AB}\approx 0$. The spin precession rate is $\omega_{0}=\frac{g\mu _{B}B}{q(0)\hbar}$.

Spin-exchange collisions preserve total angular momentum of a colliding pair of atoms but can scramble the hyperfine state of the atoms. Atoms in different hyperfine states do not precess coherently and thereby limit the coherence lifetime of the atoms. However, decoherence due to spin-exchange collisions can be nearly eliminated if the spin-exchange collisions occur much faster than the precession frequency of the atoms. In this regime of fast spin-exchange, all atoms in an ensemble rapidly change hyperfine states, spending the same amounts of time in each hyperfine state and causing the spin ensemble to precess more slowly but remain coherent\cite{Happer1977}. In the limit of fast spin-exchange and small magnetic field, the spin-exchange relaxation rate vanishes for sufficiently small magnetic field\cite{Allred2002}. In equation (9), we can find that the fundamental sensitivity of the hybrid optical pumping SERF atomic magnetometer increases when part or all of $R_{wall}^{B}$, $R_{SD}$, $R_{B}$, $R_{SE}^{BB}$, $R_{SE}^{AB}$ and $R_{SE}^{BA}$ (the later three terms are approximately zero in sufficiently low magnetic field and the magnetometer is in the SERF regime, which is helpful for us to study how $B$ influence the SERF regime and fundamental sensitivity of the magnetometer) decrease, $n_{B}$, $V$ and $t$ increase. For the expressions of $R_{wall}^{B}$, $R_{SD}$, $R_{B}$, $R_{SE}^{BB}$, $R_{SE}^{AB}$, $R_{SE}^{BA}$, $n_{B}$, $V$ and $t$, we just need to consider the fundamental sensitivity of the magnetometer change with one of the cell effective radius $a$, $n_{He}$, $n_{N2}$, $t$, cell temperature $T$, pumping rate of pump beam ($R_{p}^{K}$ and $R_{p}^{Cs}$), external magnetic field $B$ and measurement volume $V$.
Diffusion of alkali metal atoms A and B to the wall will corrode the vapor cell and decrease the fundamental sensitivity of the magnetometer. Sufficiently many buffer gas will reduce diffusion of alkali metal atoms A and B to the wall.
The probed alkali-metal atoms have a large absorption effect on the pumping beam, it's an additional relaxation item for the alkali-metal atoms of the hybrid optical pumping SERF magnetometer.
The spin exchange rate between alkali metal atoms A and alkali metal atoms B play a similar ``pumping beam'' action. Atom B is polarized by the spin exchange collision between alkali metal atoms A and B.
The pumping effect of probe beam means circularly polarized light in the probe beam pumps alkali-metal atoms.
The outer electrons of the alkali metal atoms are polarized by the pumping beam, the polarized electrons undergo Larmor precession under the external magnetic field.

If we consider the influence of the light shift noise\cite{Savukov2005,Seltzer2008} $B_{LS}$, photon shot noise\cite{Ledbetter2008} $B_{psn}$, spin-projection noise\cite{Seltzer2008} $B_{spn}$, magnetic field noise\cite{Kornack2007,Kubo1966} $B_{mag}$, technology noise $B_{tech}$ and other noise $B_{other}$ on the SERF atomic magnetometer. Using the method of superposition of power spectral density, we can obtain the sensitivity of the hybrid optical pumping SERF atomic magnetometer as following
\begin{equation}
Sen=\sqrt{\left( \delta B\right)
^{2}+B_{LS}^{2}+B_{psn}^{2}+B_{spn}^{2}+B_{mag}^{2}+B_{tech}^{2}+B_{other}^{2}%
}\text{.}  \label{9}
\end{equation}

If the noises above are optimized by technology means, the sensitivity of the hybrid optical pumping SERF atomic magnetometer approaches to the ultimate sensitivity, which is also helpful to study the atomic spin gyroscope\cite{Quan2014,Quan2014C,Quan2016,Quan2017,Quan2017O}.

We take one of the cell effective radius $a$, $n_{He}$, $n_{N2}$, $t$, cell temperature $T$, pumping rate of pump beam ($R_{p}^{K}$ and $R_{p}^{Cs}$), $B$ (it is helpful for us to study how $B$ influence the SERF regime and fundamental sensitivity of the magnetometer) and measurement volume $V$ in equation (9) as a variable (other parameters are invariable) to obtain the results that the fundamental sensitivity of the hybrid optical pumping SERF magnetometer based $^{\text{39}}$K-$^{\text{85}}$Rb-$^{\text{4}}$He and $^{\text{133}}$Cs-$^{\text{85}}$Rb-$^{\text{4}}$He vary with the variable. Depending on suggestions and the typical conditions of the experiment group\cite{Fang2014,Quan2015,Chen2016S}, in order to facilitate the theoretical analysis, we take $n_{He}=10^{19}$ $cm^{-3}$, $n_{N2}=2\times 10^{17}$ $cm^{-3}$, $T=457.5$ $K$, $R_{p}^{K}=R_{p}^{Cs}=200000$ $s^{-1}$, $a=1$ $cm$, $V=1$ $cm^{3}$, $t=100$ $s$, $B=10^{-15}$ $T$ ($R_{SE}^{ee}\approx 0$). Because $n_{K}$, $n_{Rb}$ and $n_{Cs}$ vary with $T$, the relation between the fundamental sensitivity of $^{\text{39}}$K ($^{\text{133}}$Cs)-$^{\text{85}}$Rb-$^{\text{4}}$He magnetometers and the number density of alkali-metal atoms ($n_{K}$, $n_{Rb}$ or $n_{Cs}$) is nonlinear. At the same $T$, $n_{K}$, $n_{Rb}$ and $n_{Cs} $ are different, the fundamental sensitivity are also different. We will study the vapor cell by the characteristics and properties of the microcavities\cite{QiRan2009,JiAnChun2009,JiAnChun2007}.

Figure 3 shows the relaxation rates due to diffusion of $^{\text{85}}$Rb in the $^{\text{4}}$He gas to the wall of $^{\text{39}}$K-$^{\text{85}}$Rb-$^{\text{4}}$He magnetometer $R_{wall-K-Rb-He}^{Rb-He}$, the total relaxation rates due to diffusion of $^{\text{85}}$Rb in the $^{\text{4}}$He and N$_{\text{2}}$ gas to the wall of $^{\text{39}}$K-$^{\text{85}}$Rb-$^{\text{4}}$He magnetometer $R_{wall-K-Rb-He}^{Rb-He-N2}$, the relaxation rates due to diffusion of $^{\text{85}}$Rb in the $^{\text{4}}$He gas to the wall of $^{\text{133}}$Cs-$^{\text{85}}$Rb-$^{\text{4}}$He magnetometer $R_{wall-Cs-Rb-He}^{Rb-He}$ and the total relaxation rates due to diffusion of $^{\text{85}}$Rb in the $^{\text{4}}$He and N$_{\text{2}}$ gas to the wall of $^{\text{133}}$Cs-$^{\text{85}}$Rb-$^{\text{4}}$He magnetometer $R_{wall-Cs-Rb-He}^{Rb-He-N2}$ decrease when $^{\text{4}}$He atom number density $n_{He}$ increases, the relaxation rates due to diffusion of $^{\text{85}}$Rb in the N$_{\text{2}}$ gas to the wall of $^{\text{39}}$K-$^{\text{85}}$Rb-$^{\text{4}}$He magnetometer $R_{wall-K-Rb-He}^{Rb-N2}$ and the relaxation rates due to diffusion of $^{\text{85}}$Rb in the N$_{\text{2}}$ gas to the wall of $^{\text{133}}$Cs-$^{\text{85}}$Rb-$^{\text{4}}$He magnetometer $R_{wall-Cs-Rb-He}^{Rb-N2}$ increase slowly when $n_{He}$ increases in Fig. 3(a). $R_{wall-K-Rb-He}^{Rb-N2}$, $R_{wall-K-Rb-He}^{Rb-He-N2}$, $R_{wall-Cs-Rb-He}^{Rb-N2}$ and $R_{wall-Cs-Rb-He}^{Rb-He-N2}$ decrease when $n_{N2}$ increases, $R_{wall-K-Rb-He}^{Rb-He}$ and $R_{wall-Cs-Rb-He}^{Rb-He}$ increase slowly when $n_{N2}$ increases in Fig. 3(b). $R_{wall-K-Rb-He}^{Rb-He}$, $R_{wall-K-Rb-He}^{Rb-N2}$, $R_{wall-K-Rb-He}^{Rb-He-N2}$, $R_{wall-Cs-Rb-He}^{Rb-He}$, $R_{wall-Cs-Rb-He}^{Rb-N2}$ and $R_{wall-Cs-Rb-He}^{Rb-He-N2}$ decreases slowly when pumping rate of pump beam increases in Fig. 3(c). $R_{wall-K-Rb-He}^{Rb-N2}$, $R_{wall-K-Rb-He}^{Rb-He-N2}$, $R_{wall-Cs-Rb-He}^{Rb-N2}$ and $R_{wall-Cs-Rb-He}^{Rb-He-N2}$ increase, $R_{wall-K-Rb-He}^{Rb-He}$ and $R_{wall-Cs-Rb-He}^{Rb-He}$ increase slowly when cell temperature increases in Fig. 3(d). $R_{wall-K-Rb-He}^{Rb-He}$, $R_{wall-K-Rb-He}^{Rb-N2}$, $R_{wall-K-Rb-He}^{Rb-He-N2}$, $R_{wall-Cs-Rb-He}^{Rb-He}$, $R_{wall-Cs-Rb-He}^{Rb-N2}$ and $R_{wall-Cs-Rb-He}^{Rb-He-N2}$ decreases rapidly when the cell effective radius increases in Fig. 3(e). The $R_{wall}$ of $^{\text{39}}$K-$^{\text{85}}$Rb-$^{\text{4}}$He magnetometer is bigger than the one of $^{\text{133}}$Cs-$^{\text{85}}$Rb-$^{\text{4}}$He magnetometer.

Figures 4 and 5 show the fundamental sensitivity of $^{\text{39}}$K ($^{\text{133}}$Cs)-$^{\text{85}}$Rb-$^{\text{4}}$He magnetometers as a function of $^{\text{39}}$K, $^{\text{133}}$Cs, $^{\text{85}}$Rb, $^{\text{4}}$He and N$_{\text{2}}$ number density, $R_{p}^{K}$ and $R_{p}^{Cs}$, external magnetic field, cell temperature and measurement time, cell effective radius, measurement volume, includes the effects of spin exchange due to $^{\text{39}}$K-$^{\text{85}}$Rb, $^{\text{85}}$Rb-$^{\text{85}}$Rb, $^{\text{85}}$Rb-$^{\text{39}}$K, $^{\text{133}}$Cs-$^{\text{85}}$Rb, $^{\text{85}}$Rb-$^{\text{133}}$Cs and spin relaxation due to $^{\text{39}}$K-$^{\text{85}}$Rb, $^{\text{85}}$Rb-$^{\text{85}}$Rb, $^{\text{85}}$Rb-$^{\text{39}}$K, $^{\text{133}}$Cs-$^{\text{85}}$Rb, $^{\text{85}}$Rb-$^{\text{133}}$Cs, $^{\text{85}}$Rb-$^{\text{4}}$He collisions, $^{\text{85}}$Rb-N$_{\text{2}}$ destructions.

Fig. 4(a) represents that the fundamental sensitivity of $^{\text{39}}$K-$^{\text{85}}$Rb-$^{\text{4}}$He magnetometer (black line in squares) increases with the increasing $^{\text{4}}$He number density when $^{\text{4}}$He number density is smaller than a critical value about $4.22\times 10^{19}$ $cm^{-3}$, it decreases when $^{\text{4}}$He number density is higher than the value. The fundamental sensitivity of $^{\text{133}}$Cs-$^{\text{85}} $Rb-$^{\text{4}}$He magnetometer (red line in dots) increases with the increasing $^{\text{4}}$He number density when $^{\text{4}}$He number density is smaller than a critical value about $4.15\times 10^{19}$ $cm^{-3}$ and it decreases when $^{\text{4}}$He number density is higher than the value. For this phenomenon, we think that more alkali-metal atoms diffuse to the cell wall and less spin exchange collisions between alkali-metal atoms A and B when $^{\text{4}}$He number density is smaller than the value and decrease. Less alkali-metal atoms diffuse to the cell wall and more spin exchange collisions between alkali-metal atoms and buffer gas so that there are less spin exchange collisions in alkali-metal atoms when the number density of $^{\text{4}}$He is bigger than the value and increase. Therefore, if we take the critical value as $^{\text{4}}$He number density, spin exchange collisions in alkali-metal atoms are the most, we can obtain the highest fundamental sensitivity of the magnetometer.

Fig. 4(b) shows that the fundamental sensitivity of $^{\text{39}}$K-$^{\text{85}}$Rb-$^{\text{4}}$He magnetometer (black line in squares) increases with the increasing N$_{\text{2}}$ number density when N$_{\text{2}}$ number density is smaller than a critical value about $1.22\times 10^{19}$ $cm^{-3}$, it decreases when N$_{\text{2}}$ number density is higher than the value. The fundamental sensitivity of $^{\text{133}}$Cs-$^{\text{85}} $Rb-$^{\text{4}}$He magnetometer (red line in dots) increases with the increasing N$_{\text{2}}$ number density when N$_{\text{2}}$ number density is smaller than a critical value about $1.21\times 10^{19}$ $cm^{-3}$, it decreases when N$_{\text{2}}$ number density is higher than the value. Therefore, if we take the critical value as N$_{\text{2}}$ number density, we can obtain the highest fundamental sensitivity of the magnetometer. The fundamental sensitivity of $^{\text{39}}$K ($^{\text{133}}$Cs)-$^{\text{85}} $Rb-$^{\text{4}}$He magnetometers decrease with the increasing pumping rate of pump beam respectively in Fig. 4(c). When pumping rate of pump beam is bigger than about $1916 s^{-1}$ and N$_{\text{2}}$ number density is bigger than about $1.974\times 10^{16}$ $cm^{-3}$ , the fundamental sensitivity of $^{\text{39}}$K-$^{\text{85}}$Rb-$^{\text{4}}$He magnetometer is higher than the one of $^{\text{133}}$Cs-$^{\text{85}}$Rb-$^{\text{4}}$He magnetometer. The fundamental sensitivity of $^{\text{39}}$K ($^{\text{133}}$Cs)-$^{\text{85}}$Rb-$^{\text{4}}$He magnetometers (black line in squares and red line in dots) increase with the increasing measurement time. The fundamental sensitivity of $^{\text{133}}$Cs-$^{\text{85}}$Rb-$^{\text{4}}$He magnetometer is lower than the one of $^{\text{39}} $K-$^{\text{85}}$Rb-$^{\text{4}}$He magnetometer in Fig. 4(d).

Fig. 5(a) describes that when the external magnetic field $B$ is smaller than about $10^{-8}$ $T$, the fundamental sensitivity of $^{\text{39}}$K-$^{\text{85}}$Rb-$^{\text{4}}$He magnetometer is higher than the one of $^{\text{133}}$Cs-$^{\text{85}}$Rb-$^{\text{4}}$He magnetometer and they almost do not vary with the increasing external magnetic field respectively, the fundamental sensitivity decreases rapidly when $B$ is bigger than about $10^{-8}$ $T$ and the fundamental sensitivity of $^{\text{39}}$K-$^{\text{85}}$Rb-$^{\text{4}}$He magnetometer is lower than the one of $^{\text{133}}$Cs-$^{\text{85}}$Rb-$^{\text{4}}$He magnetometer when $B$ is bigger than about $2.845\times10^{-8}$ $T$. The fundamental sensitivity of $^{\text{39}}$K ($^{\text{133}}$Cs)-$^{\text{85}} $Rb-$^{\text{4}}$He magnetometers increase with the increasing cell temperature respectively and there are more spin exchange collisions in alkali-metal atoms in Fig. 5(b). The fundamental sensitivity of $^{\text{39}}$K ($^{\text{133}}$Cs)-$^{\text{85}} $Rb-$^{\text{4}}$He magnetometers increase with the increasing cell effective radius respectively in Fig. 5(c). The fundamental sensitivity of $^{\text{39}}$K ($^{\text{133}}$Cs)-$^{\text{85}}$Rb-$^{\text{4}}$He magnetometers with $a=5 cm$ increase with increasing measurement volume respectively in Fig. 5(d). The fundamental sensitivity of $^{\text{133}}$Cs-$^{\text{85}}$Rb-$^{\text{4}}$He magnetometer is lower than the one of $^{\text{39}}$K-$^{\text{85}}$Rb-$^{\text{4}}$He magnetometer in Fig. 5(a)-(d).

As a result, the polarization of $^{\text{85}}$Rb atom of the hybrid optical pumping SERF magnetometer based on $^{\text{133}}$Cs-$^{\text{85}}$Rb-$^{\text{4}}$He is bigger than the one based on $^{\text{39}}$K-$^{\text{85}}$Rb-$^{\text{4}}$He in Fig. 2. However, the fundamental sensitivity of $^{\text{39}}$K-$^{\text{85}}$Rb-$^{\text{4}}$He magnetometer is higher than the one of $^{\text{133}}$Cs-$^{\text{85}}$Rb-$^{\text{4}}$He magnetometer when pumping rate of pump beam is bigger than about $1916 s^{-1}$ in figures 4 and 5. For another buffer gas $^{\text{21}}$Ne, a large $^{\text{85}}$Rb magnetization field due to spin interaction between $^{\text{85}}$Rb atom and $^{\text{21}}$Ne atoms causes a large spin exchange relaxation rate of $^{\text{85}}$Rb atom\cite{ChenY2016} and $^{\text{85}}$Rb atom can make $^{\text{21}}$Ne atoms hyperpolarized, which will affect the magnetic field measurement, it is a better choice to take $^{\text{4}}$He as the buffer gas of the SERF magnetometer to measure the magnetic field and take $^{\text{21}}$Ne as the buffer gas of the SERF magnetometer to measure inertia. The fundamental sensitivity of the magnetometers based on $^{\text{133}}$Cs-$^{\text{85}}$Rb-$^{\text{4}}$He is lower than the one based on $^{\text{39}}$K-$^{\text{85}}$Rb-$^{\text{4}}$He when the pumping rate of pump beam is bigger than about $1916 s^{-1}$ and $B$ is bigger than about $2.845\times10^{-8}$ $T$ (the magnetometers is not in the SERF regime for this external magnetic field).

The polarization of $^{\text{85}}$Rb atom of the magnetometer is uniform when $n_{He}=10^{19}$ $cm^{-3}$, $n_{N2}=2\times 10^{17}$ $cm^{-3}$, $T=457.5$ $K$, $R_{p}^{K}=R_{p}^{Cs}=200000$ $s^{-1}$, $a=1$ $cm$, $V=1$ $cm^{3}$, $B=10^{-15}$ $T$ and $t=100$ $s$. Under above condition, we obtain a fundamental sensitivity of $1.5073$ $aT/Hz^{1/2}$ with $^{\text{39}}$K-$^{\text{85}}$Rb-$^{\text{4}}$He magnetometer with $^{\text{85}}$Rb polarization is $0.1116$ and $n_{K}/n_{Rb}=0.0023$, a fundamental sensitivity of $1.6949$ $aT/Hz^{1/2}$ with $^{\text{133}}$Cs-$^{\text{85}}$Rb-$^{\text{4}}$He magnetometer with $^{\text{85}}$Rb polarization is $0.1864$ and $n_{Cs}/n_{Rb}=0.0151$. By optimizing above parameters, we obtain a fundamental sensitivity of $1.8359\times10^{-2}$ $aT/Hz^{1/2}$ with $^{\text{39}}$K-$^{\text{85}}$Rb-$^{\text{4}}$He with the polarization of $^{\text{85}}$Rb atom is $1.3174\times 10^{-4}$ and $n_{K}/n_{Rb}=0.0023$, a fundamental sensitivity of $1.8181\times10^{-2}$ $aT/Hz^{1/2}$ with $^{\text{133}}$Cs-$^{\text{85}}$Rb-$^{\text{4}}$He with the polarization of $^{\text{85}}$Rb atom is $2.3316\times 10^{-4}$ and $n_{Cs}/n_{Rb}=0.0151$ with $n_{He}=10^{19}$ $cm^{-3}$, $n_{N2}=2\times 10^{17}$ $cm^{-3}$, $T=457.5$ $K$, $R_{p}^{K}=R_{p}^{Cs}=200$ $s^{-1}$, $a=10$ $cm$, $B=10^{-15}$ $T$, $V=1000$ $cm^{3}$, $t=100$ $s$, with higher fundamental sensitivity possible at bigger measurement volume, proper amount of buffer gas and quench gas, smaller pumping rate of pump beam, higher temperature and longer measurement time.

\section*{Discussion}

In conclusion, we find that $^{\text{85}}$Rb polarization increases with the increasing pumping rate of pump beam. The $^{\text{85}}$Rb polarization of $^{\text{133}}$Cs-$^{\text{85}}$Rb-$^{\text{4}}$He magnetometer is bigger than the one of $^{\text{39}}$K-$^{\text{85}}$Rb-$^{\text{4}}$He magnetometer. The polarization of $^{\text{85}}$Rb atom of $^{\text{39}}$K ($^{\text{133}}$Cs)-$^{\text{85}}$Rb-$^{\text{4}}$He magnetometers almost do not vary when the number density of $^{\text{4}}$He and N$_{\text{2}}$ increase and the number density of $^{\text{4}}$He and N$_{\text{2}}$ are smaller than some critical values and decrease rapidly when the number density of buffer gas and quench gas are bigger than the values respectively. The fundamental sensitivity increases with the increasing number density of buffer gas and quench gas when the number density of buffer gas and quench gas are smaller than corresponding critical values respectively and decreases when the number density of buffer gas and quench gas are bigger than the values. The fundamental sensitivity increases with the increasing cell effective radius, measurement volume, cell temperature and measurement time respectively. The fundamental sensitivity of the magnetometers decrease with increasing $R_{p}^{K}$ and $R_{p}^{Cs}$. At the same cell temperature, the polarization of $^{\text{85}}$Rb atom of $^{\text{133}}$Cs-$^{\text{85}}$Rb-$^{\text{4}}$He magnetometer is bigger than the one of $^{\text{39}}$K-$^{\text{85}}$Rb-$^{\text{4}}$He magnetometer and the fundamental sensitivity of $^{\text{133}}$Cs-$^{\text{85}}$Rb-$^{\text{4}}$He magnetometer is lower than the one of $^{\text{39}}$K-$^{\text{85}} $Rb-$^{\text{4}}$He magnetometer when $B$ is smaller than about $2.845\times10^{-8}$ $T$ and the pumping rate of pump beam is bigger than about $1916 s^{-1}$.

From the formula of the relative velocity, $R_{wall}$, $R_{SE}^{ee}$, $n_{A}$, $n_{B}$ and equation (1), we can find that increasing the cell temperature will increase $R_{wall}$, $R_{SD}$ and the number density of alkali-metal atoms, reduce $R_{SE}^{ee}$. In general, $R_{SE}^{ee}$ is smaller than $R_{wall}$ and $R_{SD}$, raising the cell temperature resulting in an increase in the fundamental sensitivity is mainly due to the great improvement of the probed alkali-metal atomic number density when the cell temperature inceases, which has a greater influence on the fundamental sensitivity than $R_{wall}$ and $R_{SD}$. If the number density of alkali-metal atoms and cell volume are fixed, in other words, when the alkali-metal atoms in the vapor cell are fully in the vapor regime, if we continue to raise the cell temperature, the alkali-metal atoms number density will not change, $R_{SE}^{ee}$ will decrease, $R_{wall}$ and $R_{SD}$ will increase. What's more, the decreased value of $R_{SE}^{ee}$ is smaller than the increased value of $R_{wall}$ and $R_{SD}$, which will decrease the fundamental sensitivity. For example, there are certain amount of alkali-metal atoms $^{\text{39}}$K($^{\text{133}}$Cs) and $^{\text{85}}$Rb with the mole fraction of atom $^{\text{85}}$Rb is $0.97$ in the vapor cell, when $T=418.3$ $K$, all of the alkali-metal atoms become vapor, $n_{K}$=$3.0072\times 10^{11}$ $cm^{-3}$, $n_{Rb}$=$1.7385\times 10^{14}$ $cm^{-3}$, $n_{Cs}$=$2.3741\times 10^{12}$ $cm^{-3}$ with $a=1$ $cm$, if we continue to increase $T$, we will find that $n_{K}$, $n_{Cs}$ and $n_{Rb}$ do not change. When $n_{He}=10^{19}$ $cm^{-3}$, $n_{N2}=2\times 10^{17}$ $cm^{-3}$, $R_{p}^{K}=R_{p}^{Cs}=200000$ $s^{-1}$, $a=1$ $cm$, $V=1$ $cm^{3}$, $B=10^{-15}$ $T$, $t=100$ $s$, the fundamental sensitivity of $^{\text{39}}$K-$^{\text{85}} $Rb-$^{\text{4}}$He and $^{\text{133}}$Cs-$^{\text{85}} $Rb-$^{\text{4}}$He magnetometers are $3.0191$ $aT/Hz^{1/2}$ and $3.1630$ $aT/Hz^{1/2}$ at $T=418.3$ $K$, $2.4204$ $aT/Hz^{1/2}$ and $2.6163$ $aT/Hz^{1/2}$ at $T=430$ $K$.

In practical applications, we should consider some questions, one very essential question is the minimum total number of atoms necessary for the operation of the magnetometer with the desired accuracy, as well as the geometric size of the setup - how small can it be made? How does the fundamental sensitivity of the elaborated setup depend on the number of atoms?

Firstly, we can find that when number density of alkali-metal atom (which is determined by the mole fraction and the cell temperature for the hybrid vapor cell with two types of alkali-metal atoms), buffer gas and quench gas are certain, if we also know the effective radius of vapor cell, pumping rate of pump beam, external magnetic field, measurement volume and measurement time, we can obtain the corresponding total number of the atoms and the fundamental sensitivity of the magnetometer. Because the number density of alkali-metal atom is determined by the cell temperature and the mole fraction for the hybrid vapor cell with two types of alkali-metal atoms, we can find that how the fundamental sensitivity of the magnetometer depend on the number density of alkali-metal atom, buffer gas and quench gas from Fig. 4(a), Fig. 4(b) and Fig. 5(b) (the cell temperature and the mole fraction corresponds to the number density of alkali metal atoms). The less number density of buffer gas and quench gas, the bigger $R_{wall}$, it's hard to say the minimum number density of buffer gas and quench gas, but we find that there are critical values for the number density of buffer gas and quench gas to make the fundamental sensitivity of the magnetometers highest.

Secondly, the smallest geometric size of the setup and volume of vapor cell depends on the processing method and materials. For example, Griffith et al. studied a miniature atomic magnetometer integrated with flux concentrators, the magnetometer uses a millimeter scale $^{\text{87}}$Rb vapor cell ($3\times 2\times 1$ $mm^{3}$) and either mu-metal or Mn-Zn ferrite flux concentrators. They found that the minimum separation of the concentrators is limited to $2$ $mm$ by the external dimensions of the vapor cell\cite{Griffith2009} and reached a sensitivity of $10^4$ $aT/Hz^{1/2}$.

Thirdly, if the amount of $^{\text{39}}$K$ (^{\text{133}}$Cs) is little (all of $^{\text{39}}$K$ (^{\text{133}}$Cs) atoms become vapor when $T=418.3$ $K$) and there is enough $^{\text{85}}$Rb in the vapor cell, when $T$ is bigger than $418.3$ $K$. We continue to increase $T$, we will find that $n_{K}/n_{Rb}$ ($n_{Cs}/n_{Rb}$) gets bigger and bigger, the fundamental sensitivity becomes higher and higher. For instance, Fang et al.\cite{Fang2014} obtained a sensitivity of approximately $5\times 10^{3}$ $aT/Hz^{1/2}$ by optimizing the parameters of SERF magnetometer based on K-Rb hybrid optical pumping when the mole fraction of K atoms is approximately $0.03$. Ito et al. studied optimal densities of alkali metal atoms in an optically pumped K-Rb hybrid atomic magnetometer considering the spatial distribution of spin polarization, calculated the spatial distribution of the spin polarization and found that the optimal density of K atoms is $3\times10^{13}$ $cm^{-3}$ and the optimal density ratio is $n_{K}/n_{Rb}\thicksim 400$ (Rb as pump atoms and K as probe atoms) to maximize the output signal and enhance spatial homogeneity of the sensor property\cite{Ito2016}.

Fourthly, the alkali-metal atoms in the vapor cell are operated in the ``hot-gas'' regime. If the atomic gas is cooled to the state of a Bose-Einstein condensation (BEC), the operation may be essentially improved. For example, Wildermuth et al. experimentally sensed electric and magnetic fields with BEC and found this field sensor simultaneously features high spatial resolution and high field sensitivity, reached a sensitivity of $\thicksim 10^{9}$ $aT$ at $3$ $\mu m$ spatial resolution\cite{Wildermuth2006}. Therefore, we can use BEC magnetometer to obtain higher sensitivity of magnetic field and spatial resolution, which is very important for the application of the magnetometer in the field of biomedicine.

If we take $\sigma_{SD}^{K-Rb}=\sigma _{SD}^{Rb-K}\approx \sigma _{SD}^{Rb}=1.6\times 10^{-17}$ $cm^{2}$, $\sigma_{SD}^{Cs-Rb}=\sigma _{SD}^{Rb-Cs}\approx \sigma _{SD}^{Cs}=2\times 10^{-16}$ $cm^{2}$, the polarization and fundamental sensitivity of the magnetometer will decrease slightly, but it will not affect the change rule of the polarization and fundamental sensitivity discussed above. To obtain a higher fundamental sensitivity between $^{\text{39}}$K-$^{\text{85}}$Rb-$^{\text{4}}$He and $^{\text{133}}$Cs-$^{\text{85}}$Rb-$^{\text{4}}$He magnetometers, it is better to choose $^{\text{39}}$K-$^{\text{85}}$Rb-$^{\text{4}}$He magnetometer (when $B$ is smaller than about $2.845\times10^{-8}$ $T$, pumping rate of pump beam is bigger than about $1916 s^{-1}$ and N$_{\text{2}}$ number density is higher than about $1.974\times 10^{16}$ $cm^{-3}$), with $^{\text{4}}$He as the buffer gas and take the critical value of $^{\text{4}}$He number density and quench gas, increase $a$, $V$, $T$ (when the quantity of alkali-metal atoms are enough), $t$, then reduce $B$ and the pumping rate of pump beam based on actual demand of the fundamental sensitivity and spatial resolution. We estimate the fundamental sensitivity limit of the hybrid optical pumping SERF magnetometer due to the shot-noise superior to $1.8359\times10^{-2}$ $aT/Hz^{1/2}$, which is higher than the shot-noise-limited sensitivity of $1$ $aT/Hz^{1/2}$ of K SERF atomic magnetometer. We could choose suitable conditions on the basis of the experiment requirements to gain a higher sensitivity of the SERF magnetometer, keep the costs down and carry forward the miniaturization and practical application of the hybrid optical pumping SERF atomic magnetometers.

\section*{Methods}
\subsection{The fundamental sensitivity calculation details.}

We obtain the above calculation results by MATLAB and chose some special points to plot with Origin 8. The fundamental sensitivity of the hybrid optical pumping SERF atomic magnetometer was obtained by equation (9) and relevant parameters used listed in Table 1 and taking one of the cell effective radius $a$, $n_{He}$, $n_{N2}$, measurement time $t$, cell temperature $T$, pumping rate of pump beam ($R_{p}^{K}$ and $R_{p}^{Cs}$), external magnetic field $B$ and measurement volume $V$ by equation (9) as a variable (other parameters are invariable) and the fundamental sensitivity of $^{\text{39}}$K-$^{\text{85}}$Rb-$^{\text{4}}$He and $^{\text{133}}$Cs-$^{\text{85}}$Rb-$^{\text{4}}$He magnetometer vary with the variable in Fig. 4(a)-(d) and Fig. 5(a)-(d), where $n_{He}=10^{19}$ $cm^{-3}$, $n_{N2}=2\times 10^{17}$ $cm^{-3}$, $T=457.5$ $K$, $R_{p}^{K}=R_{p}^{Cs}=200000$ $s^{-1}$, $a=1$ $cm$, $V=1$ $cm^{3}$, $t=100$ $s$, $B=10^{-15}$ $T$ and the magnetometer polarization is obtained by equation (6) and its relevant parameters are chosen as above. For $R_{SE}^{KK}$, $R_{SE}^{RbRb}$, $R_{SE}^{KRb}$, $R_{SE}^{RbK}$ and their total spin-exchange collisions relaxation rate $R_{SE}^{KRbHe}$ increase, $R_{SE}^{CsCs}$, $R_{SE}^{RbRb}$, $R_{SE}^{CsRb}$, $R_{SE}^{RbCs}$ and their total spin-exchange collisions relaxation rate $R_{SE}^{CsRbHe}$ increase when $B$ increases and decrease when $T$ increases in Fig. 6.
We find that $R_{SE}^{AA}$ ($R_{SE}^{AA}=\left( \frac{g\mu _{B}B}{q(0)_{A}\hbar }\right) ^{2}\frac{q(0)_{A}^{2}-(2I_{A}+1)^{2}}{2k_{SE}^{A-A}n_{A}}$), $R_{SE}^{BB}$, $R_{SE}^{AB}$, $R_{SE}^{BA}$ and $R_{SE}^{ee}$ ($R_{SE}^{ee}=R_{SE}^{AA}+R_{SE}^{BB}+ R_{SE}^{AB}+R_{SE}^{BA}$) increase when $B$ increases and decrease when $T$ increases. When $T$ is higher than $418.3$ $K$, $B$ is smaller than $10^{-10}$ $T$, $R_{SE}^{ee}\approx 0$. When $B$ is bigger than $10^{-9}$ $T$ and $T$ is lower than $400$ $K$, we can not ignore the effect of the alkali-alkali spin-exchange collisions relaxation rate.
Therefore, there need to reduce the external magnetic field to $10^{-10}$ $T$ below and make the cell temperature higher than $418.3$ $K$ to reduce the effect of alkali-alkali spin-exchange collisions relaxation rate on the SERF regime and weak magnetic field measurement in the experiments.

\renewcommand{\baselinestretch}{1.0}
\begin{table}
\renewcommand\arraystretch{1.0}
\caption{Parameters used for the calculation}
\vspace{2.0mm}
\centering
{\tabcolsep0.5in
\begin{tabular}{cccccccccc}
  \hline
$parameter$                                                     & $value$ \\
\cline{1-2}
  Boltzmann's constant $k_{B}$                                                       & $1.38\times 10^{-23}$ $J/K$ \\
Atomic mass unit $m$                                            & $1.660539040(20)\times 10^{-27}$ $kg$ \\
$\pi$                                                           & $3.14$\\
Electron spin $g$ factor                                        & $2\times 1.001159657$ \\
Planck's constant $\hbar$                                                         & $1.054589\times 10^{-34}$ $Js$\\
Bohr magneton $\mu_B$                                           & $9.27408\times 10^{-24}$ $J/T$\\
The mole fraction of $^{\text{39}}$K                            & $0.03$\\
The mole fraction of $^{\text{133}}$Cs                          & $0.03$\\
$D_{0}\ (K-He)$\cite{Franz1982}                                 & $0.35$ $cm^{2}/s$   \\
$D_{0}\ (Rb-He)$\cite{Franz1976}                                & $0.5$ $cm^{2}/s$    \\
$D_{0}\ (Cs-He)$\cite{Franz1974}                                & $0.291$ $cm^{2}/s$   \\
$D_{0}\ (K-N2)$\cite{Silver1984}                                & $0.2$ $cm^{2}/s$   \\
$D_{0}\ (Rb-N2)$\cite{Franz1976}                                & $0.19$ $cm^{2}/s$    \\
$D_{0}\ (Cs-N2)$\cite{Franz1974}                                & $0.098$ $cm^{2}/s$   \\
$\sigma _{SE}^{K}$\cite{Aleksandrov1999}                        & $1.8\times 10^{-14}$ $cm^{2}$  \\
$\sigma_{SE}^{Rb}$\cite{Ressler1969}                            & $1.9\times 10^{-14}$ $cm^{2}$  \\
$\sigma _{SE}^{Cs}$\cite{Ressler1969}                           & $2.1\times10^{-14}$ $cm^{2}$   \\
$\sigma _{SD}^{K}$\cite{Vliegen2001}                            & $1\times 10^{-18}$ $cm^{2}$     \\
$\sigma_{SD}^{Rb}$\cite{Walker1997}                             & $1.6\times 10^{-17}$ $cm^{2}$  \\
$\sigma _{SD}^{Cs}$\cite{Bhaskar1980}                           & $2\times10^{-16}$ $cm^{2}$    \\
$\sigma _{SE}^{K-Rb}$                      & $\sigma _{SE}^{K}\frac{n_{K}}{n_{K}+n_{Rb}}+\sigma _{SE}^{Rb}\frac{n_{Rb}}{n_{K}+n_{Rb}}$\\
$\sigma _{SE}^{Rb-K}$                      & $\sigma _{SE}^{Rb}\frac{n_{Rb}}{n_{K}+n_{Rb}}+\sigma _{SE}^{K}\frac{n_{K}}{n_{K}+n_{Rb}}$ \\
$\sigma _{SE}^{Cs-Rb}$                    & $\sigma _{SE}^{Cs}\frac{n_{Cs}}{n_{Cs}+n_{Rb}}+\sigma _{SE}^{Rb}\frac{n_{Rb}}{n_{Cs}+n_{Rb}}$ \\
$\sigma _{SE}^{Rb-Cs}$                    & $\sigma _{SE}^{Rb}\frac{n_{Rb}}{n_{Cs}+n_{Rb}}+\sigma _{SE}^{Cs}\frac{n_{Cs}}{n_{Cs}+n_{Rb}}$ \\
$\sigma _{SE}^{^{\text{87}}Rb-^{\text{133}}Cs}$\cite{Gibbs1967} & $(2.3\pm 0.2)\times10^{-14}$ $cm^{2}$ \\
$\sigma _{SD}^{K-Rb}$                     & $\sigma _{SD}^{K}\frac{n_{K}}{n_{K}+n_{Rb}}+\sigma _{SD}^{Rb}\frac{n_{Rb}}{n_{K}+n_{Rb}}$ \\
$\sigma _{SD}^{Rb-K}$                     & $\sigma _{SD}^{Rb}\frac{n_{Rb}}{n_{K}+n_{Rb}}+\sigma _{SD}^{K}\frac{n_{K}}{n_{K}+n_{Rb}}$ \\
$\sigma_{SD}^{Cs-Rb}$                     & $\sigma _{SD}^{Cs}\frac{n_{Cs}}{n_{Cs}+n_{Rb}}+\sigma _{SD}^{Rb}\frac{n_{Rb}}{n_{Cs}+n_{Rb}}$ \\
$\sigma_{SD}^{Rb-Cs}$                     & $\sigma _{SD}^{Rb}\frac{n_{Rb}}{n_{Cs}+n_{Rb}}+\sigma _{SD}^{Cs}\frac{n_{Cs}}{n_{Cs}+n_{Rb}}$ \\
$\sigma _{SD}^{K-He}$\cite{Allred2002}                          & $8\times 10^{-25}$ $cm^{2}$    \\
$\sigma _{SD}^{Rb-He}$\cite{Allred2002}                         & $9\times10^{-24}$ $cm^{2}$      \\
$\sigma _{SD}^{Cs-He}$\cite{Allred2002}                         & $2.8\times 10^{-23}$ $cm^{2}$   \\
$\sigma _{SD}^{K-N_{2}}$\cite{Seltzer2008,Kadlecek1998}         & $7.9\times10^{-23}$ $cm^{2}$    \\
$\sigma _{SD}^{Rb-N_{2}}$\cite{Seltzer2008,Allred2002}          & $1\times 10^{-22}$ $cm^{2}$     \\
$\sigma _{SD}^{Cs-N_{2}}$\cite{Seltzer2008,Allred2002}          & $5.5\times 10^{-22}$ $cm^{2}$   \\
Nuclear Spin of $^{\text{39}}$K $I_{K}$                         & $1.5$  \\
Nuclear Spin of $^{\text{85}}$Rb $I_{Rb}$                       & $2.5$  \\
Nuclear Spin of $^{\text{133}}$Cs $I_{Cs}$                      & $3.5$  \\
  \hline
\end{tabular}
}
\end{table}

\begin{addendum}
\item [Acknowledgement]
We thank P. S. He (Beijing Technology and Business University), Y. Chen (Harvard University), T. Wang (University of California, Berkeley), J. X. Lu (Beihang University), H.Y. Wang (Institute of Physics, Chinese Academy of Sciences), Z. X. Deng (Sun Yat-sen University) and B. Dong (NTSC, Chinese Academy of Sciences) for discussions and communications. This work was supported by the NKRDP under grants Nos. 2016YFA0301500, NSFC under grants Nos. 61405003, 11434015, 61227902, 61378017, KZ201610005011, SKLQOQOD under grants No. KF201403, SPRPCAS under grants No. XDB01020300, XDB21030300.

\item [Author Contributions]
J.H.L. and W.M.L. proposed the ideas. J.H.L. interpreted physics, performed the theoretical as well as the numerical calculations and wrote the main manuscript. D.Y.J., L.L.W., Y.L., W.Q., J.C.F. and W.M.L. checked the calculations and the results. All of the authors reviewed the manuscript.

\item [Competing Interests]
The authors declare that they have no competing financial interests.

\item [Correspondence]
Correspondence and requests for materials should be addressed to Wu-Ming Liu (email: wliu@iphy.ac.cn).

\end{addendum}

\clearpage

\newpage \bigskip
\begin{figure}
\begin{center}
\epsfig{file=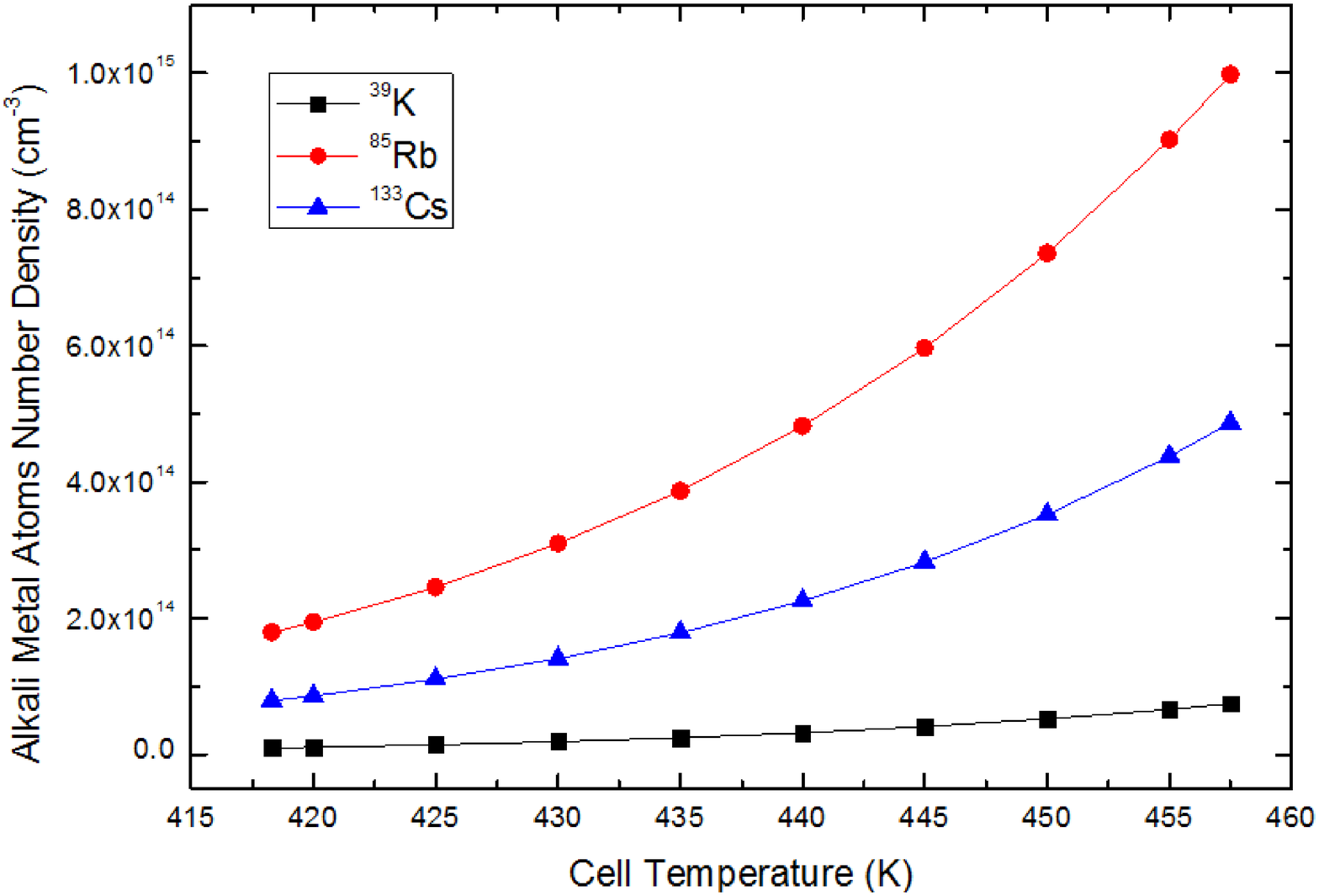,width=10cm}
\end{center}
\label{fig:number density-alkali atom and T}
\end{figure}
\textbf{Figure 1 $\arrowvert$ The number density of $^{\text{39}}$K, $^{\text{133}}$Cs and $^{\text{85}}$Rb vary with temperature.} Among $^{\text{39}}$K, $^{\text{85}}$Rb and $^{\text{133}}$Cs SERF magnetometers, there is a maximum temperature range for $^{\text{39}}$K to make the magnetometer in the SERF regime with the number density of $^{\text{39}}$K (black line in squares) satisfies the conditions of the SERF regime, so the SERF magnetometer based on $^{\text{39}}$K is suitable for an environment with the temperature varying drastically.
\bigskip
\begin{figure}
\begin{center}
\epsfig{file=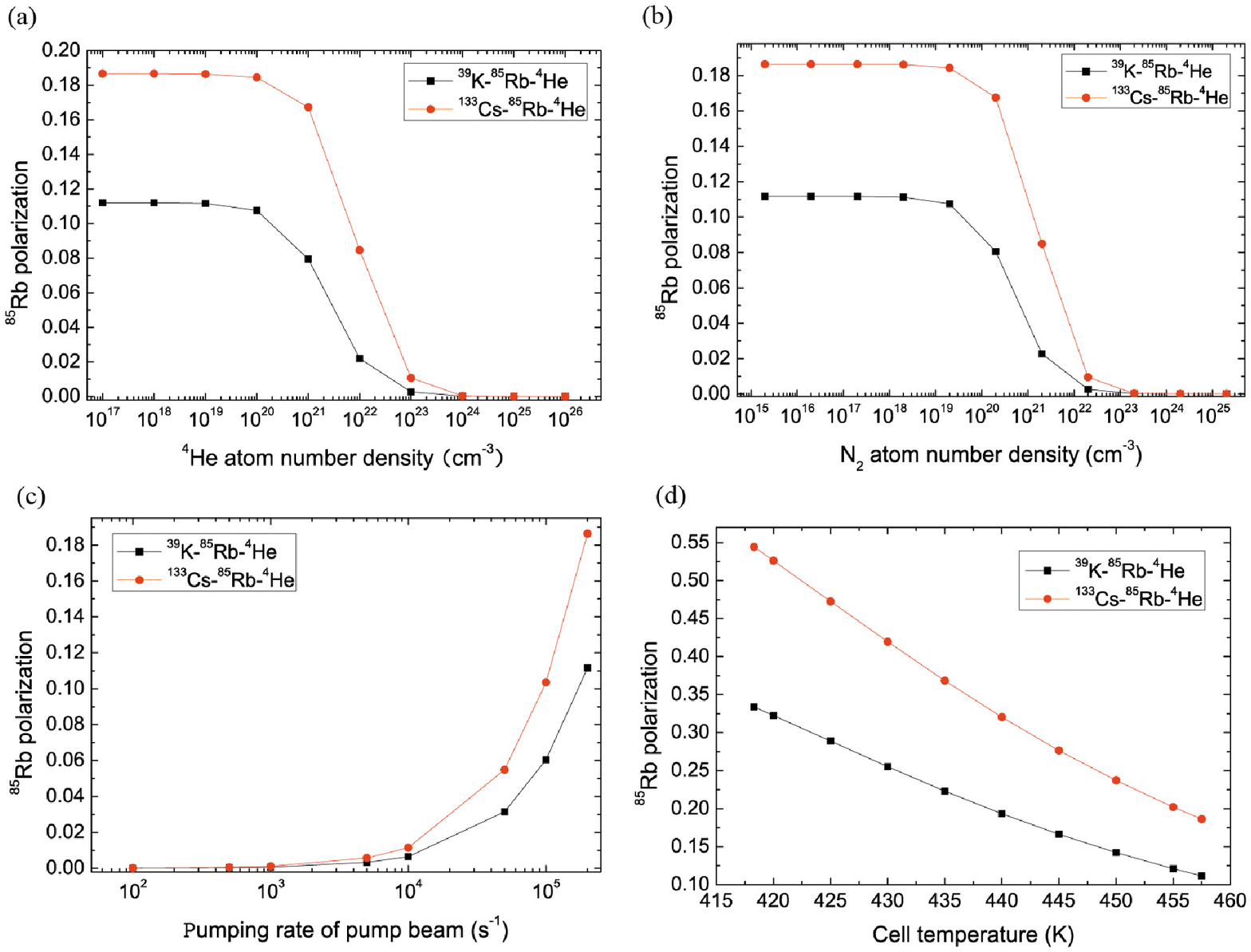, width=15cm}
\end{center}
\label{fig: polarizition,number density,optical pumping rate}
\end{figure}
\textbf{Figure 2 $\arrowvert$ The $^{\text{85}}$Rb polarization of $^{\text{39}}$K ($^{\text{133}} $Cs)-$^{\text{85}}$Rb-$^{\text{4}}$He magnetometers varies with the number density of the buffer gas ($^{\text{4}}$He), the pumping rate of pump beam, number density of the quench gas N$_{\text{\textbf{\textbf{2}}}}$ and the cell temperature respectively.} (a) $^{\text{85}}$Rb polarization almost does not vary with the increasing $^{\text{4}}$He number density when $^{\text{4}}$He number density is smaller than a critical value about $10^{20}$ $cm^{-3}$ in $^{\text{39}}$K -$^{\text{85}}$Rb-$^{\text{4}}$He (black line in squares) and $^{\text{133}}$Cs-$^{\text{85}}$Rb-$^{\text{4}}$He (red line in dots) magnetometers, otherwise, $^{\text{85}}$Rb polarization decreases rapidly. (b) $^{\text{85}}$Rb polarization almost does not vary with increasing N$_{\text{2}}$ number density when N$_{\text{2}}$ number density is smaller than about $2\times 10^{19}$ $cm^{-3}$ in $^{\text{39}}$K-$^{\text{85}}$Rb-$^{\text{4}}$He and $^{\text{133}}$Cs-$^{\text{85}}$Rb-$^{\text{4}}$He magnetometers, otherwise, the $^{\text{85}}$Rb polarization decreases rapidly. (c) $^{\text{85}}$Rb polarization increases with increasing $R_{p}^{K}$ and $R_{p}^{Cs}$ respectively. (d) $^{\text{85}}$Rb polarization decreases with the increasing cell temperature. The $^{\text{85}}$Rb polarization of $^{\text{133}}$Cs-$^{\text{85}}$Rb-$^{\text{4}}$He magnetometer is bigger than the one of $^{\text{39}}$K-$^{\text{85}}$Rb-$^{\text{4}}$He magnetometer in (a)-(d).
\bigskip
\begin{figure}
\begin{center}
\epsfig{file=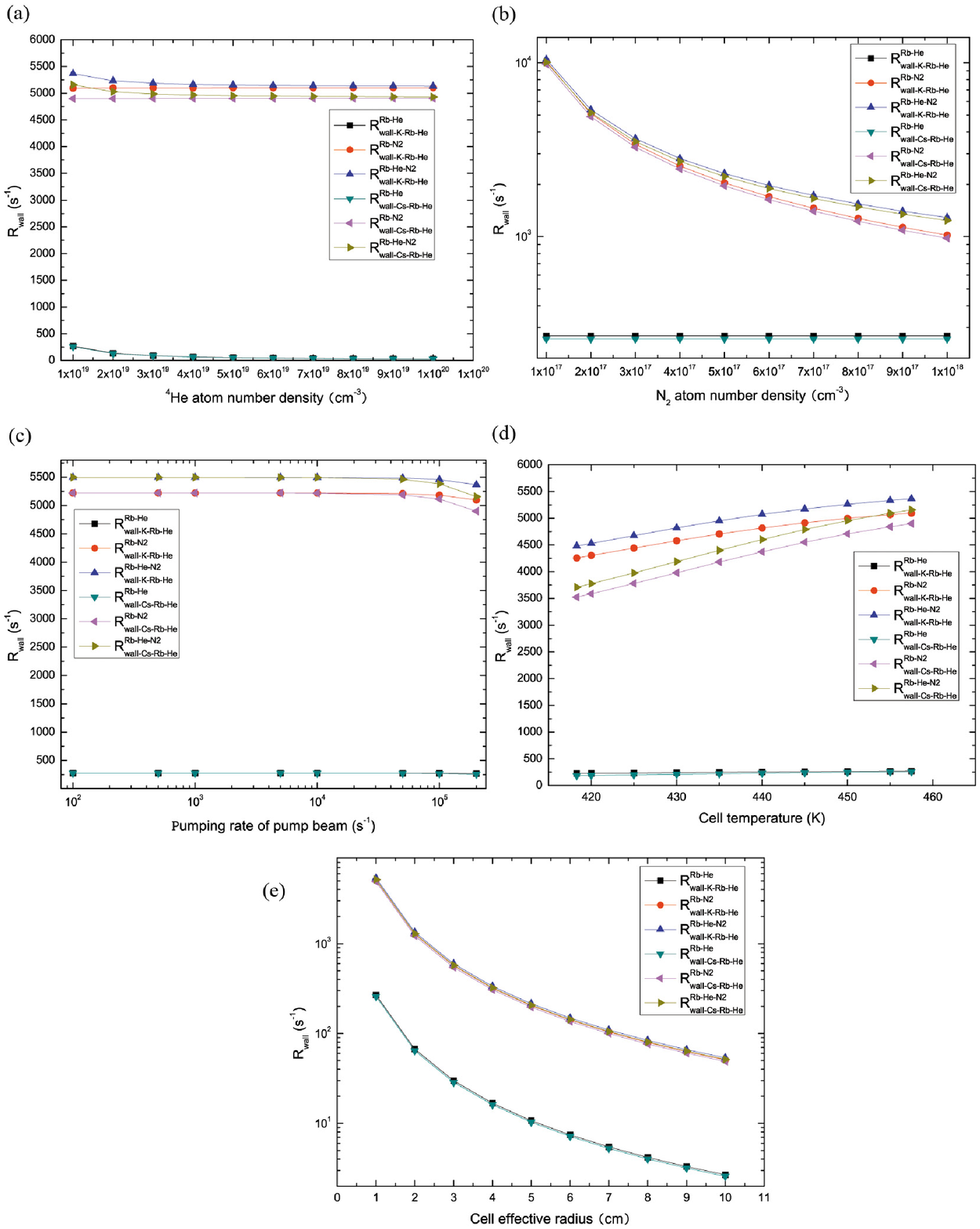,width=15cm}
\end{center}
\label{fig: Rwall,number density,optical pumping rate,temperature}
\end{figure}
\textbf{Figure 3 $\arrowvert$ The $R_{wall}$ of $^{\text{39}}$K-$^{\text{85}}$Rb-$^{\text{4}}$He and $^{\text{133}} $Cs-$^{\text{85}}$Rb-$^{\text{4}}$He magnetometers vary with the number density of buffer gas $^{\text{4}}$He and quench gas N$_{\text{2}}$, pumping rate of pump beam, cell temperature and cell effective radius.} (a) $R_{wall-K-Rb-He}^{Rb-He}$, $R_{wall-K-Rb-He}^{Rb-He-N2}$, $R_{wall-Cs-Rb-He}^{Rb-He}$ and $R_{wall-Cs-Rb-He}^{Rb-He-N2}$ decrease when $^{\text{4}}$He atom number density $n_{He}$ increases, $R_{wall-K-Rb-He}^{Rb-N2}$ and $R_{wall-Cs-Rb-He}^{Rb-N2}$ increase slowly with increasing $^{\text{4}}$He atom number density. (b) $R_{wall-K-Rb-He}^{Rb-N2}$, $R_{wall-K-Rb-He}^{Rb-He-N2}$, $R_{wall-Cs-Rb-He}^{Rb-N2}$ and $R_{wall-Cs-Rb-He}^{Rb-He-N2}$ decrease as $n_{N2}$ increases, $R_{wall-K-Rb-He}^{Rb-He}$ and $R_{wall-Cs-Rb-He}^{Rb-He}$ increase slowly when $n_{N2}$ increases. (c) $R_{wall-K-Rb-He}^{Rb-He}$, $R_{wall-K-Rb-He}^{Rb-N2}$, $R_{wall-K-Rb-He}^{Rb-He-N2}$, $R_{wall-Cs-Rb-He}^{Rb-He}$, $R_{wall-Cs-Rb-He}^{Rb-N2}$ and $R_{wall-Cs-Rb-He}^{Rb-He-N2}$ decreases slowly when pumping rate of pump beam increases. (d) $R_{wall-K-Rb-He}^{Rb-N2}$, $R_{wall-K-Rb-He}^{Rb-He-N2}$, $R_{wall-Cs-Rb-He}^{Rb-N2}$ and $R_{wall-Cs-Rb-He}^{Rb-He-N2}$ increase rapidly, $R_{wall-K-Rb-He}^{Rb-He}$ and $R_{wall-Cs-Rb-He}^{Rb-He}$ increase slowly when $T$ increases. (e) $R_{wall-K-Rb-He}^{Rb-He}$, $R_{wall-K-Rb-He}^{Rb-N2}$, $R_{wall-K-Rb-He}^{Rb-He-N2}$, $R_{wall-Cs-Rb-He}^{Rb-He}$, $R_{wall-Cs-Rb-He}^{Rb-N2}$ and $R_{wall-Cs-Rb-He}^{Rb-He-N2}$ decreases rapidly when the cell effective radius increases. The $R_{wall}$ of $^{\text{39}}$K-$^{\text{85}}$Rb-$^{\text{4}}$He magnetometer is bigger than the one of $^{\text{133}}$Cs-$^{\text{85}}$Rb-$^{\text{4}}$He magnetometer.
\bigskip
\begin{figure}
\begin{center}
\epsfig{file=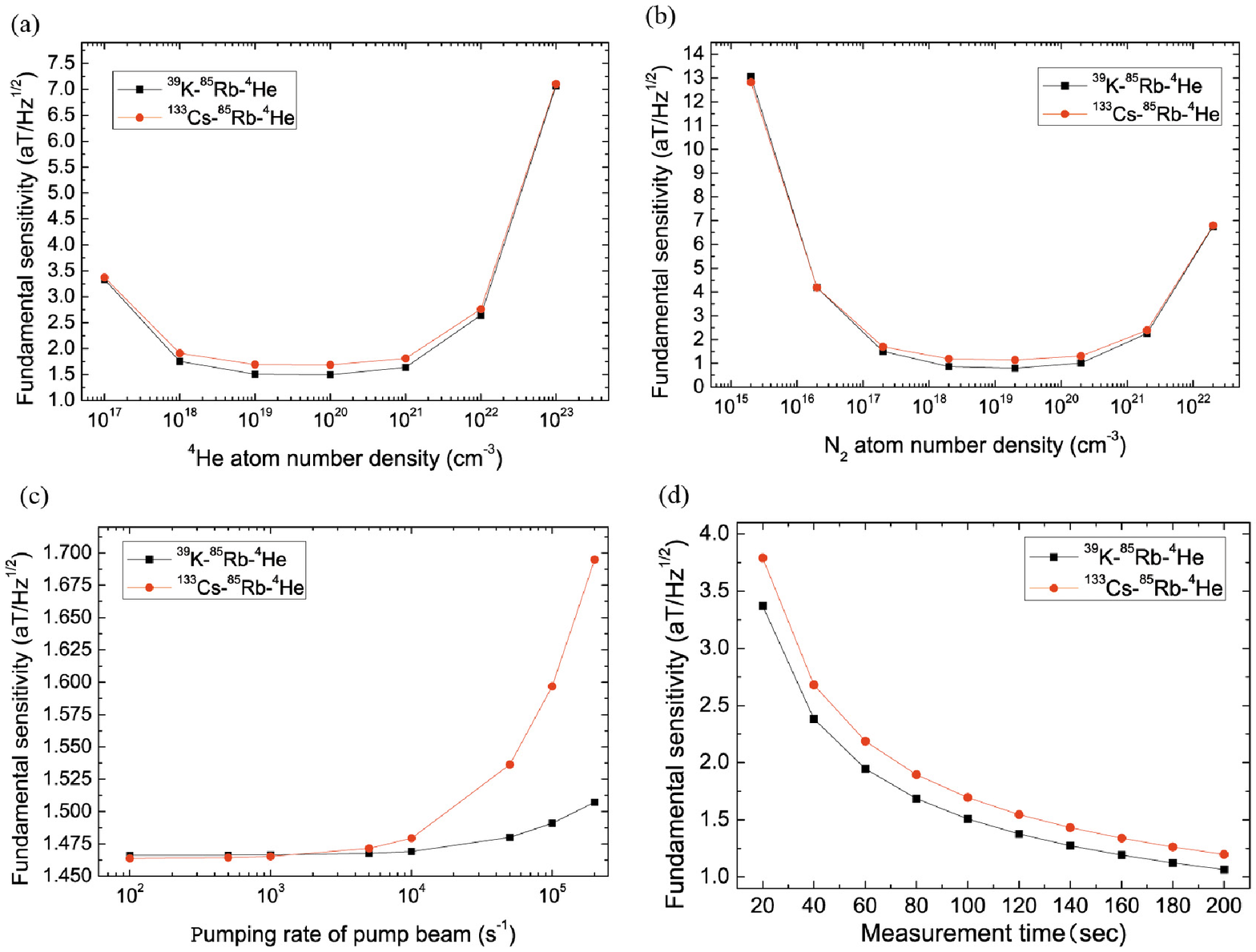,width=15cm}
\end{center}
\label{fig: sensitivity,number density}
\end{figure}
\textbf{Figure 4 $\arrowvert$ The fundamental sensitivity of $^{\text{39}}$K ($^{\text{133}}$Cs)-$^{\text{85}}$Rb-$^{\text{4}}$He magnetometers varies with the number density of buffer gas $^{\text{4}}$He and quench gas N$_{\text{2}}$, pumping rate of pump beam, measurement time.} (a) The fundamental sensitivity of $^{\text{39}}$K-$^{\text{85}} $Rb-$^{\text{4}}$He magnetometer (black line in squares) increases with the increasing number density of $^{\text{4}}$He when $^{\text{4}}$He number density is smaller than a critical value about $4.22\times 10^{19}$ $cm^{-3}$ and decreases when $^{\text{4}}$He number density is bigger than the value. The fundamental sensitivity of $^{\text{133}}$Cs-$^{\text{85}} $Rb-$^{\text{4}}$He magnetometer (red line in dots) increases with the increasing number density of $^{\text{4}}$He when $^{\text{4}}$He number density is smaller than a critical value about $4.15\times 10^{19}$ $cm^{-3}$ and decreases when $^{\text{4}}$He number density is bigger than the value. (b) The fundamental sensitivity of $^{\text{39}}$K-$^{\text{85}}$Rb-$^{\text{4}}$He magnetometer (black line in squares) increases with the increasing N$_{\text{2}}$ number density when N$_{\text{2}}$ number density is smaller than a critical value about $1.22\times 10^{19}$ $cm^{-3}$ and decreases when N$_{\text{2}}$ number density is higher than the value. The fundamental sensitivity of $^{\text{133}}$Cs-$^{\text{85}} $Rb-$^{\text{4}}$He magnetometer(red line in dots) increases with the increasing N$_{\text{2}}$ number density when N$_{\text{2}}$ number density is smaller than a critical value about $1.21\times 10^{19}$ $cm^{-3}$ and decreases when N$_{\text{2}}$ number density is higher than the value. The fundamental sensitivity of $^{\text{39}}$K ($^{\text{133}}$Cs)-$^{\text{85}} $Rb-$^{\text{4}}$He magnetometers decrease with the increasing pumping rate of pump beam in (c). The fundamental sensitivity of $^{\text{39}}$K ($^{\text{133}}$Cs)-$^{\text{85}}$Rb-$^{\text{4}}$He magnetometers increase with the increasing measurement time. The fundamental sensitivity of $^{\text{133}}$Cs-$^{\text{85}}$Rb-$^{\text{4}}$He magnetometer is lower than the one of $^{\text{39}}$K-$^{\text{85}}$Rb-$^{\text{4}}$He magnetometer in (d).
\bigskip
\begin{figure}
\begin{center}
\epsfig{file=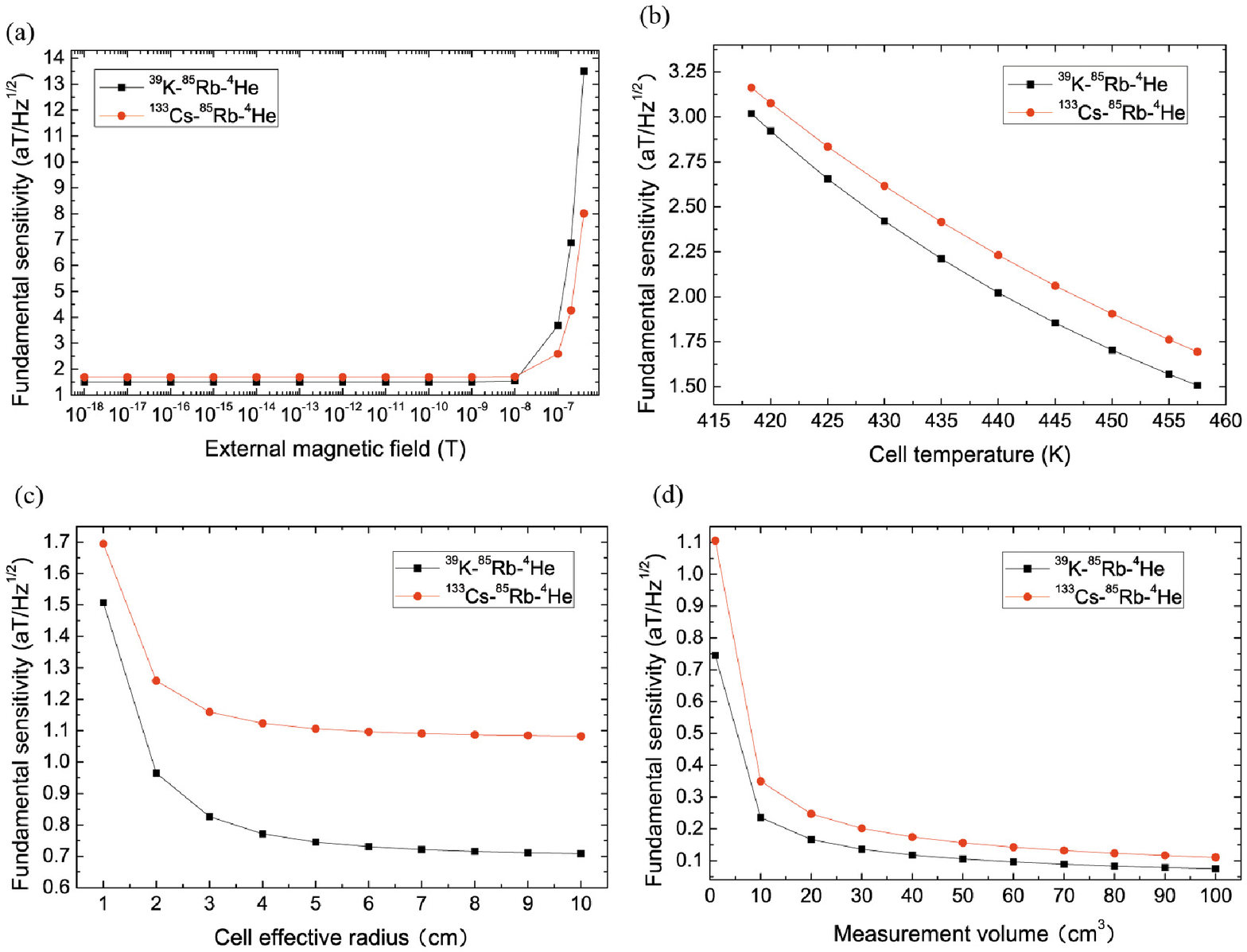,width=15cm}
\end{center}
\label{fig: sensitivity,volume}
\end{figure}
\textbf{Figure 5 $\arrowvert$ The fundamental sensitivity of $^{\text{39}}$K ($^{\text{133}}$Cs)-$^{\text{85}}$Rb-$^{\text{4}}$He magnetometers varies with the external magnetic field, cell temperature, cell effective radius and measurement volume.} (a) When the external magnetic field is smaller than about $10^{-8}$ $T$, the fundamental sensitivity of $^{\text{39}}$K ($^{\text{133}}$Cs)-$^{\text{85}} $Rb-$^{\text{4}}$He magnetometers almost do not vary with the increasing external magnetic field respectively. (b) The fundamental sensitivity of $^{\text{39}}$K ($^{\text{133}}$Cs)-$^{\text{85}} $Rb-$^{\text{4}}$He magnetometers increase with the increasing cell temperature respectively. (c) The fundamental sensitivity of $^{\text{39}}$K ($^{\text{133}}$Cs)-$^{\text{85}} $Rb-$^{\text{4}}$He magnetometers increase with the increasing cell effective radius respectively. (d) The fundamental sensitivity of $^{\text{39}}$K ($^{\text{133}}$Cs)-$^{\text{85}}$Rb-$^{\text{4}}$He magnetometers with $a=5 cm$ increase with increasing measurement volume respectively.
\bigskip
\begin{figure}
\begin{center}
\epsfig{file=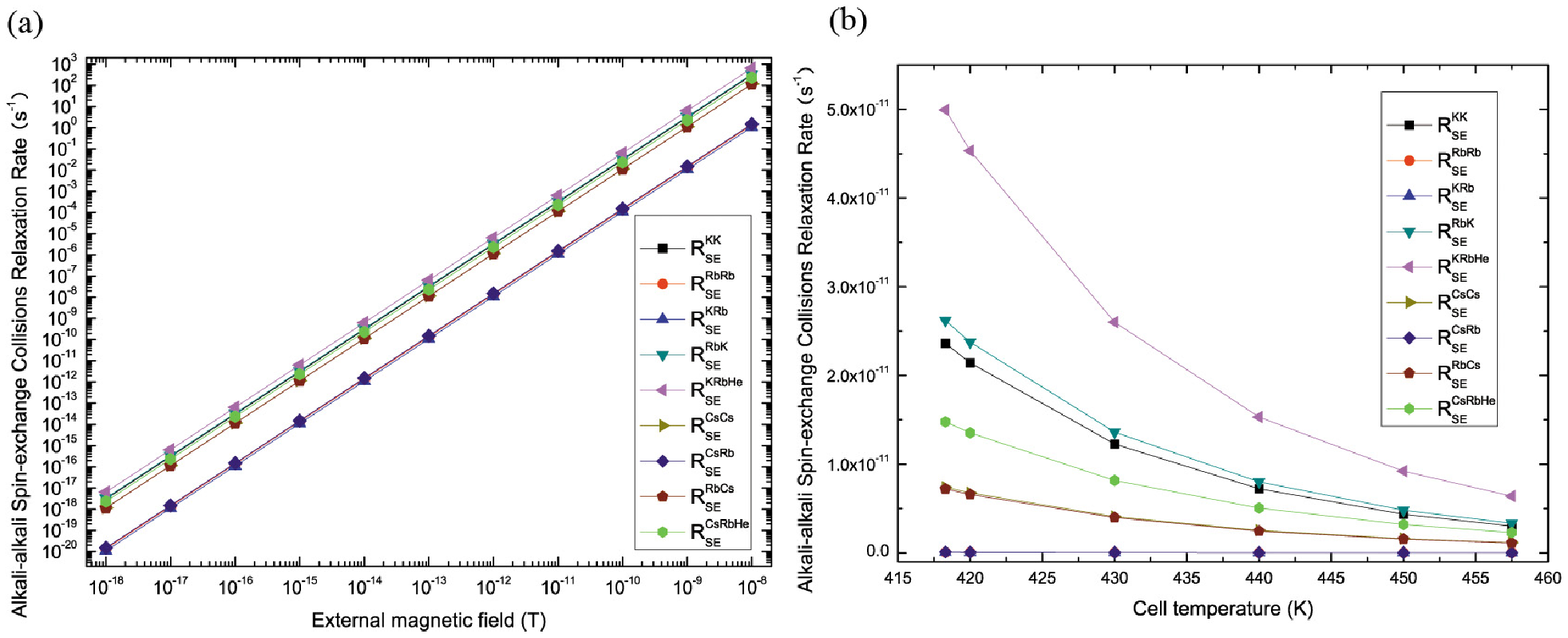,width=15cm}
\end{center}
\label{fig: Rse,B,T}
\end{figure}
\textbf{Figure 6 $\arrowvert$ The alkali-alkali spin-exchange collisions relaxation rate of $^{\text{39}}$K ($^{\text{133}}$Cs)-$^{\text{85}}$Rb-$^{\text{4}}$He magnetometers varies with the external magnetic field and cell temperature.} $R_{SE}^{KK}$, $R_{SE}^{RbRb}$, $R_{SE}^{KRb}$, $R_{SE}^{RbK}$ and their total spin-exchange collisions relaxation rate $R_{SE}^{KRbHe}$ increase, $R_{SE}^{CsCs}$, $R_{SE}^{RbRb}$, $R_{SE}^{CsRb}$, $R_{SE}^{RbCs}$ and their total spin-exchange collisions relaxation rate $R_{SE}^{CsRbHe}$ increase when $B$ increases and decrease when $T$ increases in (a) and (b).
\bigskip

\end{document}